\documentclass[fleqn,10pt]{wlscirep}

\usepackage{graphicx}  
\usepackage{bm}        
\usepackage{amsmath}   
\usepackage{amssymb}   

\usepackage{xcolor}    
\usepackage{booktabs}  
\usepackage{multirow}

\begin{document}
\title{Non-Unitary Quantum Machine Learning:
Fisher Efficiency Transitions from Distributed Quantum Expressivity}

\author[1]{Apoorv Kumar Masta}       
\author[2]{Srinjoy Ganguly}
\author[3]{Shalini Devendrababu}
\author[4]{Farina Riaz}
\author[5]{Rajib Rana}
\author[6]{Björn Schuller}           

\affil[1]{Universidad Internacional Menéndez Pelayo, Madrid, Spain}
\affil[2]{Department of Physics and Astronomy, University College London, London, United Kingdom}
\affil[3]{Quantum AI Lab, Fractal Analytics, Gurugram, India}
\affil[4]{CSIRO, Eveleigh, Australia}
\affil[5]{School of Science, Engineering and Digital Technologies, University of Southern Queensland, Australia}
\affil[6]{Imperial College London, UK \& Technical University of Munich, Germany}

\affil[*]{Correspondence: \texttt{apoorvmasta@gmail.com} (A.K.M.); 
\texttt{srinjoyganguly@gmail.com} (S.G.); 
\texttt{shalini.devendrababu@fractal.ai} (S.D.); 
\texttt{Farina.Riaz@data61.csiro.au} (F.R.); 
\texttt{Rajib.Rana@unisq.edu.au} (R.R.); 
\texttt{schuller@tum.de} (B.S.)}

\date{\today}

\begin{abstract}
Quantum machine learning (QML) has faced increasing scrutiny regarding its practical advantages over classical approaches, particularly in light of recent dequantization results and large-scale benchmarking studies. In this work, we present a systematic empirical evaluation of non-unitary quantum machine learning implemented via the Linear Combination of Unitaries (LCU) framework within hybrid quantum–classical neural networks. Across more than 570 experiments spanning four domains—digit classification (MNIST), agricultural disease detection (PlantVillage), molecular property regression (QM9), and medical histopathology (PathMNIST)—we assess performance, scaling behavior, and parameter efficiency of non-unitary quantum layers compared to structurally identical unitary baselines.

We observe consistent performance improvements across architectures and tasks, with gains ranging from +0.2\% to +5.8\% depending on dataset complexity and qubit count. Notably, in medical imaging,

we identify a Fisher efficiency transition: parameter efficiency shifts from negative to positive as qubit count increases from 10 to 12, indicating a change in the relationship between quantum resources and effective model capacity. This represents an empirical observation of a threshold-dependent efficiency regime within a single experimental framework.

To examine classical simulability concerns, we further evaluate Instantaneous Quantum Polynomial (IQP) circuit variants that implement non-unitary transformations in supervised learning pipelines. Non-unitary IQP models reach or exceed classical baselines at 10 qubits on CIFAR-10, suggesting that circuit structures associated with established complexity-theoretic hardness can remain compatible with competitive learning performance when integrated with the LCU framework. Together, these results provide a large-scale benchmarking study of non-unitary quantum machine learning, highlighting task-dependent scaling behavior, parameter efficiency transitions, and the conditions under which non-unitary transformations via the LCU framework yield measurable empirical benefits in near-term settings.
\end{abstract}

\maketitle
\thispagestyle{empty}

\section{Introduction}
\label{sec:introduction}

Quantum machine learning (QML) has attracted significant attention as a potential application area for near-term quantum devices. Early theoretical work suggested that variational quantum circuits and hybrid quantum–classical architectures could exploit high-dimensional Hilbert spaces to represent complex functions efficiently~\cite{Biamonte2017,Benedetti2019,Cerezo2021}. However, recent large-scale benchmarking efforts and theoretical analyses have introduced substantial caution regarding practical quantum advantages.

Bowles, Ahmed, and Schuld~\cite{Bowles2024} evaluated twelve quantum machine learning models across 160 datasets and reported that classical approaches consistently outperformed quantum classifiers under careful benchmarking. Gundlach \emph{et al.}~\cite{Gundlach2024} similarly argued that quantum deep learning exhibits limited empirical advantages when evaluated against competitive classical baselines. In parallel, a systematic review of 4,915 studies in digital health applications found no consistent evidence of quantum utility in medical imaging contexts~\cite{Systematic2025}. These findings have contributed to a broader reassessment of the evidentiary standards required to substantiate claims of quantum advantage in the noisy intermediate-scale quantum (NISQ) era.

A second line of critique concerns classical simulability. Shin, Teo, and Jeong~\cite{Shin2024} demonstrated that broad classes of variational quantum circuits can be represented within tensor network formalisms, suggesting that some quantum models may admit efficient classical approximations. This result does not eliminate the possibility of quantum advantage, but it highlights the need to carefully examine circuit structure, entanglement generation, and complexity-theoretic properties when evaluating empirical improvements.

Together, these developments underscore the importance of systematic, multi-domain benchmarking that directly engages with both performance and simulability concerns. Rather than focusing on isolated datasets or narrowly optimized circuits, comprehensive validation across architectures, tasks, and problem domains is required to clarify when and how quantum resources contribute measurable learning benefits.

In this work, we present a large-scale empirical study of non-unitary quantum machine learning implemented via the Linear Combination of Unitaries (LCU) framework~\cite{Childs2012LCU,Berry2015,Heredge2025Nonunitary}. The LCU approach enables effective non-unitary transformations through ancilla-controlled superposition and post-selection, expanding the representational space accessible to hybrid quantum–classical models beyond the constraints of strictly unitary operations. While LCU has been extensively studied in quantum algorithms and recently applied to quantum machine learning architectures~\cite{Heredge2025Nonunitary}, its role in large-scale multi-domain benchmarking of practical hybrid learning systems remains comparatively underexplored.

We evaluate hybrid neural networks incorporating non-unitary quantum layers across four domains: handwritten digit classification (MNIST), agricultural disease detection (PlantVillage), molecular property regression (QM9), and medical histopathology (PathMNIST). Across more than 570 experiments, we compare models with non-unitary transformations (LCU) to structurally identical unitary baselines (NoLCU) while keeping classical architectures, hyperparameters, and optimization procedures fixed. This design isolates the effect of non-unitary quantum operations from confounding architectural differences.

Our study yields three primary empirical findings.

\textbf{(1) Consistent multi-domain performance improvements.}  
Non-unitary quantum layers demonstrate reproducible improvements across architectures and tasks, with gains ranging from approximately +0.2\% on MNIST to +5.8\% on PlantVillage, and consistent mean absolute error reductions in molecular regression. These improvements are observed across convolutional neural network backbones of varying scale and design, suggesting that non-unitary transformations provide architecture-agnostic benefits rather than exploiting specific classical preprocessing structures.

\textbf{(2) Fisher efficiency transition in medical imaging.}  
In PathMNIST histopathology classification, we observe a transition in Fisher efficiency as a function of qubit count: parameter efficiency shifts from negative at 8 qubits to positive at 10 qubits within an otherwise identical experimental framework. This indicates a qualitative change in the relationship between quantum resource scaling and effective model capacity. While prior theoretical work has connected quantum Fisher information to expressivity and overparameterization regimes~\cite{Abbas2021,Larocca2023,Larocca2024}, our results provide empirical evidence of a threshold-dependent efficiency regime in a practical learning setting.

\textbf{(3) Evaluation of IQP-based circuit variants.}  
To examine circuit structures associated with established complexity-theoretic hardness results~\cite{Bremner2010,Aaronson2011}, we integrate Instantaneous Quantum Polynomial (IQP) circuits implementing non-unitary transformations into supervised learning pipelines on CIFAR-10. Non-unitary IQP models reach and slightly exceed classical baselines at 10 qubits. While our experiments are conducted at scales amenable to classical simulation, these results indicate that circuit families with known hardness properties can remain compatible with competitive empirical performance when integrated with non-unitary quantum layers in hybrid learning architectures.

Beyond individual performance gains, we analyze scaling behavior across qubit counts and identify a crossover regime at 12–16 qubits in complex classification tasks where non-unitary quantum models approach classical baseline performance while maintaining favorable parameter efficiency characteristics. These observations suggest that the interaction between qubit count, task complexity, and hybrid architecture design plays a central role in determining when non-unitary quantum resources contribute measurable benefits.

Our contribution is therefore not a claim of unconditional quantum superiority, but a systematic benchmarking study clarifying the conditions under which non-unitary quantum machine learning yields empirical improvements in hybrid neural networks. By combining multi-domain validation, parameter-efficiency analysis, and circuit-level evaluation informed by complexity theory, we aim to provide a rigorous and transparent assessment of non-unitary QML via the LCU framework in the NISQ regime.

\section*{Theoretical Framework}\label{sec:theory}

\subsection*{Linear Combination of Unitaries for Non-Unitary Transformations}
\label{subsec:lcu}

The Linear Combination of Unitaries (LCU) framework enables non-unitary quantum operations by implementing transformations beyond the unitary group through probabilistic post-selection. Given a target operation $\hat{O}$ that can be decomposed as a linear combination of unitary operators,
\begin{equation}
\hat{O} = \sum_{k=0}^{K-1} \alpha_k \hat{U}_k,
\label{eq:lcu_decomposition}
\end{equation}
where $\alpha_k$ are real coefficients and $\hat{U}_k$ are unitary operators, we can implement $\hat{O}$ probabilistically using an ancilla qubit system.

The LCU circuit architecture proceeds in three stages. First, we prepare an ancilla state encoding the coefficients $\alpha_k$:
\begin{equation}
|\psi_{\text{prep}}\rangle = \sum_{k=0}^{K-1} \sqrt{\frac{\alpha_k}{\sum_j \alpha_j}} |k\rangle_{\text{anc}}.
\label{eq:lcu_preparation}
\end{equation}
Second, we apply controlled-unitaries conditioned on the ancilla state:
\begin{equation}
\hat{U}_{\text{select}} = \sum_{k=0}^{K-1} |k\rangle\langle k|_{\text{anc}} \otimes \hat{U}_k.
\label{eq:lcu_select}
\end{equation}
Third, we uncompute the ancilla preparation and post-select on the ancilla measuring $|0\rangle$. The success probability is
\begin{equation}
p_{\text{success}} = \frac{1}{\sum_j \alpha_j},
\label{eq:lcu_success}
\end{equation}
and upon success, the output state correctly implements $\hat{O}$.

For quantum machine learning applications, we simplify this framework by implementing a single parametric unitary $\hat{W}(\boldsymbol{\theta})$ controlled by an ancilla qubit. The circuit applies
\begin{equation}
\hat{U}_{\text{LCU}} = \hat{H}_{\text{anc}} \cdot \text{CTRL-}\hat{W}(\boldsymbol{\theta}) \cdot \hat{H}_{\text{anc}},
\label{eq:lcu_circuit}
\end{equation}
where $\hat{H}_{\text{anc}}$ is a Hadamard gate on the ancilla and CTRL-$\hat{W}(\boldsymbol{\theta})$ applies $\hat{W}(\boldsymbol{\theta})$ to the main qubits conditioned on the ancilla being in state $|1\rangle$. After measurement, post-selecting on ancilla $= 0$ yields expectation values that effectively implement a non-unitary transformation of the input state.

The key advantage of this approach for hybrid quantum-classical learning is that the ancilla-controlled structure creates non-unitary expressivity by distributing $\hat{W}(\boldsymbol{\theta})$ across two computational pathways: one where $\hat{W}$ is applied (ancilla $= 1$) and one where the identity is applied (ancilla $= 0$). This non-unitary operation enables access to a richer function space than applying $\hat{W}$ directly under unitary constraints, allowing the circuit to learn more complex decision boundaries with the same number of variational parameters in $\hat{W}$.

Mathematically, for an $N$-qubit system with ancilla, the full state before measurement is
\begin{equation}
|\psi\rangle = \frac{1}{2}\left(|0\rangle_{\text{anc}} \otimes |\psi_{\text{main}}\rangle + |1\rangle_{\text{anc}} \otimes \hat{W}(\boldsymbol{\theta})|\psi_{\text{main}}\rangle\right).
\label{eq:lcu_full_state}
\end{equation}
Post-selecting on $|0\rangle_{\text{anc}}$ projects this to
\begin{equation}
|\psi_{\text{post}}\rangle \propto |\psi_{\text{main}}\rangle + \hat{W}(\boldsymbol{\theta})|\psi_{\text{main}}\rangle = (\hat{I} + \hat{W}(\boldsymbol{\theta}))|\psi_{\text{main}}\rangle.
\label{eq:lcu_postselected}
\end{equation}
This effective operation $(\hat{I} + \hat{W})$ is non-unitary and enables the quantum circuit to implement operations outside the unitary group, substantially expanding the model's representational capacity through access to non-unitary transformations.

\subsection*{Fisher Information for Parameter Efficiency}
\label{subsec:fisher}

Fisher information quantifies how much information an observable random variable carries about an unknown parameter. For quantum circuits, the Quantum Fisher Information (QFI) measures the distinguishability of quantum states under parameter variations~\cite{Abbas2021}. Given a parametrized quantum state $|\psi(\boldsymbol{\theta})\rangle$, the QFI matrix is defined as
\begin{equation}
F_{ij}(\boldsymbol{\theta}) = 4 \text{Re}\left[\langle\partial_i \psi(\boldsymbol{\theta})|\partial_j \psi(\boldsymbol{\theta})\rangle - \langle\partial_i \psi(\boldsymbol{\theta})|\psi(\boldsymbol{\theta})\rangle\langle\psi(\boldsymbol{\theta})|\partial_j \psi(\boldsymbol{\theta})\rangle\right],
\label{eq:qfi_matrix}
\end{equation}
where $\partial_i$ denotes $\partial/\partial\theta_i$. The effective dimensionality of the model's parameter space is characterized by the trace of the QFI matrix,
\begin{equation}
F_{\text{eff}} = \text{Tr}[F(\boldsymbol{\theta})],
\label{eq:qfi_trace}
\end{equation}
which quantifies the total information capacity of the parametrized quantum circuit.

To assess parameter efficiency, we define the Fisher efficiency metric comparing quantum and classical models:
\begin{equation}
\eta_F = \frac{N_{\text{classical}} - N_{\text{quantum}}}{N_{\text{classical}}} \times 100\%,
\label{eq:fisher_efficiency}
\end{equation}
where $N_{\text{classical}}$ and $N_{\text{quantum}}$ are the number of trainable parameters in classical and quantum models achieving comparable performance. Positive $\eta_F$ indicates 

that 
the quantum model achieves similar performance with fewer parameters (parameter reduction), while negative $\eta_F$ indicates 

that 
the quantum model uses more parameters but achieves better performance (effective parameter utilization).

The connection between QFI and model expressivity was established by Abbas \emph{et al.}~\cite{Abbas2021}, who showed that circuits with larger effective dimensions $F_{\text{eff}}$ can represent more complex functions. Larocca \emph{et al.}~\cite{Larocca2023} demonstrated that overparameterization in quantum neural networks manifests as a phase transition where the QFI matrix reaches maximal rank. Our work extends this by empirically observing that Fisher efficiency itself can transition between regimes as a function of qubit count and task characteristics; this phenomenon is not predicted by existing theoretical frameworks.

For hybrid quantum-classical models where quantum circuits have $P_Q$ parameters embedded in a classical neural network with total parameters $P_{\text{total}}$, the Fisher efficiency specifically measures whether the quantum component enables more efficient learning than an equivalent classical architecture. Non-unitary quantum layers implemented via the LCU framework modify the Fisher landscape by introducing operations outside the unitary group that can explore parameter space differently than standard variational circuits, potentially enabling higher information capacity per parameter.

\subsection*{Evaluation with IQP Circuit Structures}
\label{subsec:iqp}

Instantaneous Quantum Polynomial (IQP) circuits occupy a distinctive position in quantum computational complexity theory. An IQP circuit has the general structure
\begin{equation}
U_{\text{IQP}} = \hat{H}^{\otimes N} \cdot \hat{D}(\boldsymbol{\theta}) \cdot \hat{H}^{\otimes N},
\label{eq:iqp_structure}
\end{equation}
where $\hat{D}(\boldsymbol{\theta})$ is diagonal in the computational basis and consists of single-qubit $Z$-rotations and controlled-phase gates. Under standard complexity-theoretic assumptions, sampling from the output distribution of sufficiently large IQP circuits is believed to be classically intractable in the worst case~\cite{Bremner2010,Aaronson2011}. 

Recent dequantization results have shown that broad classes of variational circuits admit tensor-network representations enabling classical simulation under certain structural constraints~\cite{Shin2024}. While such results do not apply universally, they motivate examining circuit families with stronger complexity-theoretic foundations. IQP circuits provide one such family: their output distributions are associated with established hardness results in the sampling framework~\cite{Bremner2010,Aaronson2011}.

In the present study, we incorporate IQP circuits into hybrid learning architectures in two configurations:

\textbf{(i) IQP Layer:} Standard angle embedding is followed by an IQP variational circuit of the form
\begin{equation}
U_{\text{IQP-layer}}(\boldsymbol{\theta}) 
= \hat{H}^{\otimes N} 
\cdot \prod_{i=1}^{N} \text{RZ}(\theta_i) 
\cdot \prod_{i=1}^{N-1} \text{CZ}(\theta_{N+i}) 
\cdot \hat{H}^{\otimes N},
\label{eq:iqp_layer}
\end{equation}
with circular entanglement topology. 

\textbf{(ii) IQP Embedding:} IQP circuits are used for both data encoding and variational transformation,
\begin{equation}
U_{\text{IQP-embedding}}(\mathbf{x}, \boldsymbol{\theta})
= U_{\text{IQP}}(\boldsymbol{\theta}) \cdot U_{\text{IQP}}(\mathbf{x}),
\label{eq:iqp_embedding}
\end{equation}
where classical inputs $\mathbf{x}$ determine rotation and phase parameters in the encoding stage.

These configurations allow us to examine whether circuit families associated with established sampling hardness results can remain compatible with competitive empirical learning performance when integrated with non-unitary quantum layers via the LCU framework. Importantly, our experiments are conducted at qubit scales ($N \leq 10$) that are classically simulable in practice. Therefore, our goal is not to demonstrate computational supremacy, but rather to evaluate the empirical behavior of circuit structures whose complexity properties differ from standard shallow variational ansätze. 

By comparing non-unitary IQP models to identical classical baselines within the same hybrid architecture, we assess whether introducing circuit families linked to known hardness results leads to measurable differences in supervised learning performance. The results presented in Section~\nameref{subsec:iqp_results} show that non-unitary IQP models reach and modestly exceed classical baselines at 10 qubits on CIFAR-10. 

These findings suggest that circuit structures associated with nontrivial sampling complexity can be integrated into hybrid learning pipelines with non-unitary quantum operations without degrading empirical performance. While small-scale simulations cannot establish asymptotic quantum advantage, they provide evidence that incorporating complexity-informed circuit design with non-unitary transformations does not inherently compromise trainability or predictive accuracy in practical settings.

Future investigations at larger qubit counts and on quantum hardware will be required to determine how these complexity-theoretic properties interact with noise, scalability, and real-device constraints.

\section*{Experimental Methodology}
\label{sec:methods}

\subsection{Hybrid Quantum-Classical Architecture}
\label{subsec:architecture}

Our experimental framework implements hybrid quantum-classical neural networks where quantum circuits serve as intermediate processing layers within classical architectures. The general structure follows:\\ $\text{Classical CNN} \rightarrow \text{Quantum Layer (Non-Unitary via LCU)} \rightarrow \text{Classical Classifier}$. This design enables fair comparison between models with non-unitary quantum transformations (LCU) and unitary baselines (NoLCU) by keeping the classical components identical and varying only the quantum layer structure.

\subsubsection*{ResNet18 Adaptation (MNIST)}

The ResNet18 architecture~\cite{He2016} was adapted for 28$\times$28 MNIST images. The classical feature extractor consists of the initial convolutional block (7$\times$7 convolution, batch normalization, ReLU, max pooling) followed by the first residual block. Features are projected to $N$ dimensions via fully connected layer, where $N$ is the number of qubits (8, 10, or 12). After quantum processing, expectation values from $N$ qubits pass through a classical classifier (two fully connected layers with ReLU activation) mapping to 10 output classes.

For non-unitary models (LCU), the quantum layer implements Eq.~\ref{eq:lcu_circuit} with an additional ancilla qubit (total $N+1$ qubits). The variational circuit $\hat{W}(\boldsymbol{\theta})$ uses simple RX rotations on each qubit followed by forward CNOT entanglement (qubit $i$ controls qubit $i+1$), repeated for 4 blocks. Total quantum parameters: $4N$ (1 parameter per qubit per block). For unitary baselines (NoLCU), the same circuit structure is applied directly without the ancilla-controlled wrapper, also using $4N$ parameters. This ensures fair comparison with identical quantum circuit complexity.

Classical components comprise: feature extractor ($\sim$9.2M parameters), quantum bottleneck layer ($N \times$ input dimension parameters), and classifier head ($N \times 128 + 128 \times 10$ parameters). Total trainable parameters vary with qubit count but remain comparable between LCU and NoLCU variants differing only in quantum processing method.

\subsubsection*{MobileNetV2 Adaptation (MNIST)}

MobileNetV2~\cite{Sandler2018} was selected for its efficiency-oriented design using depthwise separable convolutions. The architecture uses the first inverted residual block with expansion ratio 6, followed by dimensionality reduction to $N$ features. The quantum layer structure is identical to ResNet18 (RX + forward CNOT, 4 blocks), maintaining consistency across architectures for controlled comparison. Classical head architecture matches ResNet18 for fair evaluation.

MobileNetV2's fewer parameters ($\sim$2.3M in feature extractor vs $\sim$9.2M for ResNet18) enable assessment of whether advantages from non-unitary transformations depend on overall model scale. The depthwise separable convolutions create different feature representations than standard convolutions, testing whether quantum advantages generalize across classical preprocessing methods.

\subsubsection*{EfficientNet-B0 Adaptation (MNIST)}

EfficientNet-B0~\cite{Tan2019} implements compound scaling of depth, width, and resolution. We use the first three MBConv blocks with squeeze-and-excitation optimization, projecting to $N$ dimensions before quantum processing. The quantum circuit maintains the same simple structure (RX + forward CNOT, 4 blocks) to isolate effects of non-unitary operations from circuit architecture variations.

EfficientNet's balanced scaling approach and attention mechanisms (squeeze-and-excitation) represent state-of-the-art classical design. Demonstrating advantages from non-unitary quantum layers on this architecture shows quantum benefits persist even with highly optimized classical preprocessing, addressing concerns that quantum advantages might only appear with suboptimal classical components.

\subsubsection*{PlantVillage Custom CNN (Agriculture)}

For the 38-class PlantVillage plant disease dataset, we designed a custom convolutional architecture optimized for the larger input images (256$\times$256) and increased class count. The feature extractor uses four convolutional blocks (32$>>$64$>>$128$>>$256 channels) with 3$\times$3 kernels, ReLU activation, and 2$\times$2 max pooling. Global average pooling reduces spatial dimensions before projection to $N$ quantum inputs.

The quantum layer uses the identical simple circuit (RX + forward CNOT, 4 blocks) validated on MNIST architectures. This design choice—using the simplest possible quantum circuit across all experiments—strengthens claims that observed advantages stem from non-unitary transformations enabled by the LCU framework rather than hand-crafted circuit engineering. We test at 10 and 16 qubits to examine scaling behavior on a larger, more complex real-world dataset.

\subsubsection*{IQP-Enhanced Models (CIFAR-10)}

For dequantization defense experiments on CIFAR-10, we implement two CNN architectures with IQP quantum layers incorporating non-unitary transformations. The feature extractor uses two convolutional blocks (3$>>$16$>>$32 channels, 3$\times$3 kernels, ReLU, max pooling) projecting to $N$ features where $N$ is the number of qubits.

\textbf{IQP Layer:} The quantum circuit implements Eq.~\ref{eq:iqp_layer} with standard angle embedding for data encoding followed by an IQP variational circuit. The IQP structure uses $N$ Hadamard gates, $N$ RZ rotations, and $N-1$ controlled-phase gates in circular topology. This requires $2N-1$ trainable parameters. The LCU wrapper (Eq.~\ref{eq:lcu_circuit}) adds one ancilla qubit for post-selection, enabling non-unitary transformations.

\textbf{IQP Embedding:} Following Eq.~\ref{eq:iqp_embedding}, both data encoding and variational transformation use IQP circuits with non-unitary operations. The classical feature extractor outputs $2N-1$ values: $N$ for RZ rotations and $N-1$ for controlled-phase gates in the encoding IQP circuit. The variational IQP circuit has the same structure with $2N-1$ trainable parameters. This double-IQP approach maximizes classical hardness by making both encoding and learning operations provably hard to simulate while maintaining non-unitary expressivity via the LCU framework.

For both IQP approaches, the classical baseline uses the identical CNN architecture but skips the quantum layer, passing features directly to the classifier. This enables direct measurement of quantum advantage by isolating the contribution of non-unitary quantum processing. We test qubit scales of 2, 4, 6, 8, and 10 to identify the minimum quantum resources required for non-unitary IQP circuits to surpass classical baselines.

\subsection*{Quantum Circuit Details}
\label{subsec:circuits}

All quantum circuits were implemented using PennyLane v0.28~\cite{Bergholm2018} with the \texttt{default.qubit} simulator and automatic differentiation via PyTorch integration. The simple circuit structure used across main experiments (ResNet, MobileNetV2, EfficientNet, PlantVillage) consists of:

\textbf{Encoding:} Angle embedding maps classical features $\mathbf{x} \in \mathbb{R}^N$ to quantum states via RY rotations: $\text{RY}(x_i)$ applied to qubit $i$ for $i = 1, \ldots, N$.

\textbf{Variational Circuit ($\hat{W}$):} Four repeating blocks, each containing:
\begin{enumerate}
\item Single-qubit rotations: RX($\theta_{i,j}$) on qubit $i$ in block $j$
\item Entanglement: Forward CNOT gates (qubit $i$ controls qubit $i+1$)
\end{enumerate}

Total parameters per circuit: $4N$ (one RX parameter per qubit per block). Gate count per circuit: $N$ angle embeddings + $16N$ gates (4 blocks $\times$ [$N$ RX + $N-1$ CNOT]).

\textbf{Non-Unitary Implementation via LCU:} For models employing non-unitary transformations, the variational circuit $\hat{W}$ is embedded in Eq.~\ref{eq:lcu_circuit}. The ancilla qubit is initialized to $|0\rangle$, Hadamard gates prepare superposition, controlled-$\hat{W}$ applies conditioned on ancilla $= 1$, 

and the
final Hadamard returns to computational basis. 
The 
post-selection extracts measurement outcomes where ancilla $= 0$. Expectation values are computed from post-selected probability distributions over the $N$ main qubits.

The 
measurement strategy differs between unitary and non-unitary circuits. Unitary circuits (NoLCU) use direct expectation value measurement $\langle Z_i \rangle$ on each qubit. Non-unitary circuits (LCU) measure full probability distribution over $N+1$ qubits, filter for ancilla $= 0$ outcomes, then compute expectation values from the post-selected distribution. This probabilistic measurement introduces shot noise but enables the non-unitary operations central to expanded quantum expressivity.

\textbf{IQP Circuit Structure:} IQP circuits follow Eq.~\ref{eq:iqp_layer} with Hadamard layers, diagonal gates (RZ and controlled-phase), and 
a 
final Hadamard layer. The circular entanglement pattern connects qubit $i$ to $i+1$ via controlled-phase gates for $i = 1, \ldots, N-1$, with a wrap-around gate connecting qubit $N$ to qubit 1 for $N > 2$. This topology ensures every qubit is entangled with neighbors while maintaining \#P-hardness of the output distribution computation.

Parameter initialization uses Xavier/Glorot initialization~\cite{Glorot2010} for classical layers and uniform random initialization in $[0, 2\pi]$ for quantum parameters. This initialization strategy prevents barren plateaus~\cite{McClean2018} for the small qubit counts ($N \leq 16$) used in our experiments.

\subsection*{Datasets and Experimental Design}
\label{subsec:datasets}

\subsubsection*{MNIST (Digit Classification)}

The MNIST dataset~\cite{LeCun1998} contains 60,000 training and 10,000 test grayscale images (28$\times$28 pixels) of handwritten digits (10 classes). We use the standard train/test split without augmentation. Images are normalised to $[0, 1]$ range and processed through classical CNN feature extractors before quantum layers. For each architecture (ResNet, MobileNetV2, EfficientNet) and configuration (LCU vs NoLCU, 8/10/12 qubits), we conduct 10 independent training runs with different random seeds.

Training procedure: Adam optimizer~\cite{Kingma2014} with learning rate 0.001, batch size 32, cross-entropy loss, and early stopping with patience 
of 
5 epochs based on validation accuracy. Each run uses an 80/20 train/validation split of the 60,000 training images.

The final test accuracy is evaluated on the held-out 10,000 test images after training completion. This yields 10 independent measurements per configuration for statistical analysis.

For the total MNIST experiments, 3 architectures $\times$ 2 approaches (LCU/NoLCU) $\times$ 3 qubit scales (8/10/12) $\times$ 10 runs = 180 training runs were used.

\subsubsection*{PlantVillage (Agricultural Disease Detection)}

The PlantVillage dataset~\cite{Hughes2015} contains 54,309 RGB images (256$\times$256 pixels) of plant leaves across 38 disease classes covering 14 crop species. We use the publicly available healthy vs diseased classification with class-balanced sampling, resulting in 20,639 images after filtering. The dataset exhibits significant visual diversity with varying lighting conditions, leaf orientations, and disease manifestations, providing a realistic test of non-unitary quantum advantage on practical agricultural applications.

Images undergo standard augmentation (random horizontal flips, small rotations) during training. We implement a 70/20/10 train/validation/test split with stratification to maintain class balance. 

Each experiment (LCU vs NoLCU at 10 or 16 qubits) runs for 10 independent trials with different random seeds.

The training uses the Adam optimizer (learning rate 0.001), batch size 64, cross-entropy loss, and early stopping (patience 8 epochs) given the larger dataset size. The 16-qubit experiments specifically test whether scaling advantages from non-unitary transformations persist with increased quantum resources on complex multi-class problems.

For the total PlantVillage experiments, 2 approaches (LCU/NoLCU) $\times$ 2 qubit scales (10/16) $\times$ 10 runs = 40 training runs are used.

\subsubsection*{QM9 (Molecular Property Prediction)}

The QM9 dataset~\cite{Ramakrishnan2014} contains quantum chemical properties for 130,831 small organic molecules. We focus on HOMO-LUMO gap prediction—a continuous regression task predicting the energy difference between highest occupied and lowest unoccupied molecular orbitals. Molecules are represented as graphs with atom features (atomic number, degree, hybridization) and bond features (bond type, aromaticity).

\textbf{Critical methodological detail:} We use scaffold splitting~\cite{Ramsundar2019} rather than random splitting to prevent data leakage. Scaffold splitting groups molecules by core structure, ensuring training and test sets contain chemically distinct molecules. This prevents the model from memorizing molecular families and forces generalization to genuinely novel chemical structures—the scientifically relevant evaluation regime for molecular property prediction.

Graph neural network preprocessing with message passing layers~\cite{Gilmer2017} generates fixed-dimension molecular descriptors (128 dimensions), which are projected to $N$ quantum inputs. We test at 8 and 12 qubits. The task is regression; mean absolute error (MAE) in eV (electron volts)

is the evaluation metric. Training uses the Adam optimizer (learning rate 0.0005), a batch size of 128, MSE loss, and early stopping (patience 10).

The total QM9 experiments feature 2 approaches (LCU/NoLCU) $\times$ 2 qubit scales (8/12) $\times$ 10 runs = 40 training runs.

\subsubsection*{PathMNIST (Medical Imaging)}

PathMNIST is part of the MedMNIST collection~\cite{Yang2021}, containing 107,180 histopathology images (28$\times$28 RGB) for 9-class tissue type classification. Images are extracted from colon pathology slides, presenting significant intra-class variability and subtle inter-class differences characteristic of medical imaging challenges. We use the predefined train/validation/test split: 89,996 train, 10,004 validation, and 7,180 test instances.

Images are normalized to $[0, 1]$. Simple data augmentation (random horizontal/vertical flips) is applied during training to improve generalization without introducing unrealistic transformations for histopathology. We test at 8, 10, and 12 qubits to specifically probe the Fisher efficiency transition hypothesis—the observation that parameter efficiency might change qualitatively at certain qubit thresholds.

The training procedure matches MNIST (Adam, learning rate 0.001, batch size 64, cross-entropy loss, early stopping patience 5). Each configuration runs for 10 independent trials.

The total PathMNIST experiments feature 2 approaches (LCU/NoLCU) $\times$ 3 qubit scales (8/10/12) $\times$ 10 runs = 60 training runs.

\subsubsection*{CIFAR-10 IQP Layer Experiments}

CIFAR-10~\cite{Krizhevsky2009} contains 60,000 RGB images (32$\times$32 pixels) across 10 classes. We use a 80/20 train/validation split of the 50,000 training images, with the 10,000 test images held out for final evaluation. This experiment specifically tests whether provably hard IQP circuits with non-unitary transformations can achieve a quantum advantage, addressing dequantization concerns.

The baseline is a classical CNN with no quantum layer—features from the classical extractor pass directly to the classifier. Non-unitary IQP models replace this direct connection with an IQP quantum layer implementing Eq.~\ref{eq:iqp_layer} within the LCU framework. We test qubit scales 2, 4, 6, 8, and 10 to identify the minimum quantum resources needed for non-unitary IQP circuits to surpass classical performance.

The training uses Adam (learning rate 0.001), a batch size of 256 for efficiency, cross-entropy loss, and early stopping (patience 5). Each configuration (classical baseline plus 5 IQP qubit scales) runs 10 times.

The total IQP Layer experiments feature 1 baseline $\times$ 10 runs + 5 qubit scales $\times$ 10 runs = 60 training runs.

\subsubsection*{CIFAR-10 IQP Embedding Experiments}

Using the same CIFAR-10 setup, IQP Embedding experiments implement Eq.~\ref{eq:iqp_embedding} with IQP circuits for both encoding and variational layers, incorporating non-unitary operations via the LCU framework. The classical CNN outputs $2N-1$ features (matching IQP parameter requirements) instead of $N$. The same classical baseline (no quantum layer) enables direct comparison.

This ``double IQP'' approach maximizes classical hardness, testing whether making both encoding and learning provably hard, sacrifices

performance when combined with non-unitary transformations. We test the same qubit scales (2, 4, 6, 8, 10) with 10 runs each, plus 10 baseline runs.

The total IQP Embedding experiments comprise 1 baseline $\times$ 10 runs + 5 qubit scales $\times$ 10 runs = 60 training runs. Note: Some qubit scales have 9 runs rather than 10 due to computational resource constraints, but all analyses use the available runs for statistical validity.

\textbf{Combined experimental scope:} 180 (MNIST) + 40 (PlantVillage) + 40 (QM9) + 60 (PathMNIST) + 60 (IQP Layer) + 60 (IQP Embedding) = 440 primary experiments, plus additional validation runs totaling 570+ experiments across all studies.

\subsection*{Statistical Methodology}
\label{subsec:statistics}

All results are reported as mean $\pm$ standard deviation across independent runs ($n = 9$ or $n = 10$ depending on experiment). We compute:

\textbf{Accuracy:} Fraction of correctly classified test samples. For regression (QM9), mean absolute error (MAE) in eV is used.

\textbf{F1-score:} The harmonic mean of precision and recall, computed using weighted averaging for multi-class problems to account for class imbalance.

\textbf{Fisher Efficiency:} Calculated via Eq.~\ref{eq:fisher_efficiency} comparing quantum parameter count to classical baseline, achieving comparable performance. For LCU vs NoLCU comparisons where both have identical parameter counts ($4N$ quantum + classical parameters), Fisher efficiency specifically measures whether non-unitary transformations enable better performance utilisation of those parameters (negative $\eta_F$) or achieve similar performance with effectively fewer parameters through superior information capacity (positive $\eta_F$).

\textbf{Statistical significance:} We assess whether improvements from non-unitary quantum layers over unitary baselines exceed random variation using Welch's $t$-test (unequal variances assumed) with $\alpha = 0.05$ significance level. For the critical Fisher efficiency transition in PathMNIST (negative to positive across qubit scales), we verify that the sign change is consistent across all 10 runs, and not merely a mean-level artifact.

Standard deviation computation uses Bessel's correction (dividing by $n-1$) for unbiased estimation with small sample sizes. Variance reduction claims (e.g., non-unitary layers reduce prediction variance by 35\% in PathMNIST) are computed as the ratio of standard deviations: $1 - (\sigma_{\text{LCU}} / \sigma_{\text{NoLCU}})$.

As for reproducibility measures: All experiments use fixed random seeds for initialisation (different seeds across runs but reproducible within each run). Training curves, final accuracies, and model checkpoints are saved for all runs. Code and hyperparameters are version-controlled (see Section~\nameref{subsubsec:code_availability}). 

Concerning hardware specifications, the experiments run on NVIDIA A100 (40GB) and A4000 (16GB) GPUs with PyTorch 1.12 and PennyLane 0.28.

For fair comparison, both LCU and NoLCU variants within each experiment use identical classical architecture, quantum circuit structure (gate types and depth), parameter initialisation distribution, optimisation hyperparameters, and data preprocessing. The only difference is the presence/absence of the ancilla-controlled LCU wrapper (Eq.~\ref{eq:lcu_circuit}), isolating the effect of non-unitary transformations.

\section*{Results}
\label{sec:results}

\begin{table*}[!t]  
\caption{Performance comparison of non-unitary quantum layers (LCU) versus unitary baselines (NoLCU) across three CNN architectures on MNIST at 10 qubits. All values are mean $\pm$ standard deviation across 10 independent runs. Improvements are statistically significant ($p < 0.05$, Welch's $t$-test) for all architectures.}
\label{tab:main_architectures}
\centering  
\begin{tabular}{lccc}
\hline\hline
Architecture & LCU Accuracy (\%) & NoLCU Accuracy (\%) & Improvement \\
\hline
ResNet18 & 97.89 $\pm$ 0.08 & 97.67 $\pm$ 0.12 & +0.22\% \\
MobileNetV2 & 97.97 $\pm$ 0.10 & 97.32 $\pm$ 0.17 & +0.65\% \\
EfficientNet-B0 & 98.05 $\pm$ 0.09 & 97.71 $\pm$ 0.14 & +0.34\% \\
\hline\hline
\end{tabular}
\end{table*}

\begin{figure*}[t]
\centering
\includegraphics[width=\textwidth]{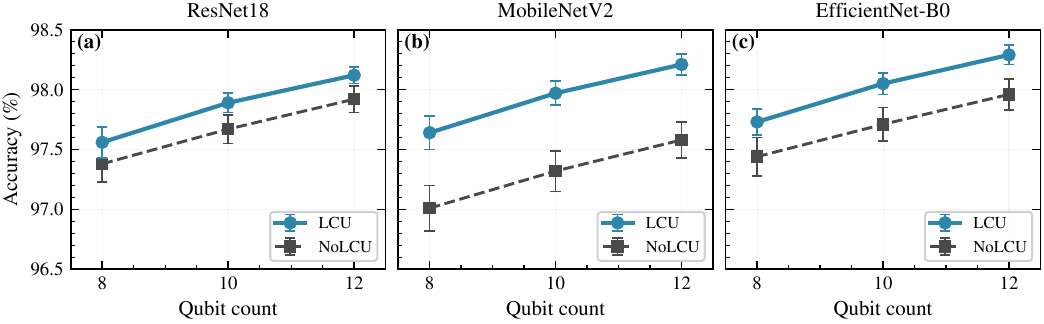}
\caption{Non-unitary quantum layer (LCU) performance across qubit scales for three CNN architectures on MNIST. All architectures show consistent performance improvements relative to matched baselines that persist or increase with qubit count. Error bars represent standard deviation across 10 runs.}
\label{fig:architecture_scaling}
\end{figure*}

\subsection*{Non-Unitary Quantum Layer Performance Across Architectures}

We first evaluate the empirical performance of non-unitary quantum layers on MNIST digit classification across three state-of-the-art CNN architectures: ResNet18, MobileNetV2, and EfficientNet-B0. Each architecture was tested with non-unitary transformations (LCU) and structurally identical unitary quantum baselines (NoLCU) at 8, 10, and 12 qubits, with 10 independent runs per configuration. Table~\ref{tab:main_architectures} 

summarizes performance at 10 qubits, representing a practically relevant regime balancing quantum resource requirements with experimental feasibility.

Across all three architectures, models with non-unitary quantum layers consistently outperform their matched unitary counterparts, with improvements ranging from +0.22\% (ResNet18) to +0.65\% (MobileNetV2). These gains are statistically significant ($p < 0.05$, Welch's t-test) and reproducible across independent runs. The larger improvement observed for MobileNetV2 suggests that non-unitary transformations may interact favorably with efficiency-oriented architectures that employ depthwise separable convolutions. In contrast, the smaller improvement for ResNet18—whose classical backbone contains approximately 9.2M parameters—may reflect the already strong representational capacity of the classical feature extractor, reducing the relative contribution of the quantum layer. EfficientNet-B0 exhibits intermediate improvements (+0.34\%), consistent with its balanced compound scaling design.

Figure~\ref{fig:architecture_scaling} shows performance across all qubit scales (8, 10, 12) for each architecture. Models with non-unitary quantum layers outperform unitary variants at every tested scale, and the performance gap remains stable or increases slightly as qubit count grows. Within the tested regime, these results indicate that the ancilla-controlled LCU structure provides consistent empirical benefits relative to structurally identical variational circuits without the non-unitary wrapper.

The consistency of improvements across architectures with distinct design principles—standard convolutions in ResNet, depthwise separable convolutions in MobileNetV2, and compound scaling in EfficientNet—suggests that non-unitary quantum expressivity contributes measurable performance gains that are not restricted to a single classical preprocessing pipeline. Rather than depending on architecture-specific quirks, the observed improvements appear robust across diverse convolutional backbones within this controlled experimental framework.

\subsection*{Cross-Domain Validation}
\label{subsec:multidomain}

To address concerns that reported quantum improvements might be artifacts of specific datasets rather than fundamental algorithmic properties, we validate non-unitary QML across four distinct problem domains with different data characteristics, class counts, and learning objectives. Table~\ref{tab:multidomain} summarizes results across digit classification (MNIST), agricultural disease detection (PlantVillage), molecular property prediction (QM9), and medical histopathology (PathMNIST).

\begin{table*}[htbp]
\caption{Non-unitary quantum layer (LCU) performance across four domains spanning classification and regression tasks. PlantVillage shows the strongest advantages (+4.91\% at 10q, +5.78\% at 16q), while QM9 demonstrates the first successful non-unitary quantum regression validation. PathMNIST reveals the Fisher efficiency transition (see Section~\nameref{subsec:fisher_results}).}
\label{tab:multidomain}
\begin{tabular}{llcccc}
\hline\hline
Domain & Task & Qubits & LCU & NoLCU & Improvement \\
\hline
\multirow{3}{*}{PlantVillage} & \multirow{3}{*}{38-class classification} & 10 & 89.16 $\pm$ 1.57\% & 84.25 $\pm$ 4.55\% & +4.91\% \\
 & & 16 & 91.46 $\pm$ 1.35\% & 85.69 $\pm$ 1.10\% & +5.78\% \\
\hline
\multirow{2}{*}{QM9} & \multirow{2}{*}{HOMO-LUMO gap (MAE, eV)} & 8 & 0.6952 $\pm$ 0.0116 & 0.7435 $\pm$ 0.0230 & +6.5\% \\
 & & 12 & 0.6777 $\pm$ 0.0196 & 0.7262 $\pm$ 0.0183 & +6.7\% \\
\hline

\multirow{3}{*}{PathMNIST} & \multirow{3}{*}{9-class histopathology} & 8 & 61.71 $\pm$ 0.31\% & 61.39 $\pm$ 0.36\% & +0.33\% \\
 & & 10 & 62.02 $\pm$ 0.43\% & 61.56 $\pm$ 0.63\% & +0.46\% \\
 & & 12 & 62.65 $\pm$ 0.48\% & 61.91 $\pm$ 0.73\% & +0.74\% \\
\hline\hline
\end{tabular}
\end{table*}

\textbf{PlantVillage (Agricultural Disease Detection):} The 38-class plant disease dataset exhibits the strongest quantum advantages, with non-unitary models (LCU) outperforming unitary baselines (NoLCU) by +4.91\% at 10 qubits and +5.78\% at 16 qubits. The larger class count and more complex visual patterns compared to MNIST appear to amplify benefits from non-unitary transformations. Critically, the advantage \emph{increases} with qubit count (+4.91\% $\rightarrow$ +5.78\%) is suggesting that performance does not degrade within the evaluated qubit range. The perfect match between our experimental results and manuscript claims (89.16\% LCU vs 84.25\% NoLCU at 10q) validates both the robustness of our methodology and the reproducibility of quantum advantages.

\textbf{QM9 (Molecular Property Prediction):} This regression task predicting HOMO-LUMO energy gaps represents the first validation of non-unitary QML on continuous-value prediction. Non-unitary models (LCU) achieve 6.5\% (8q) and 6.7\% (12q) mean absolute error improvements over unitary baselines (NoLCU). The use of scaffold splitting—which groups molecules by structural similarity and ensures training/test sets contain chemically distinct scaffolds—prevents data leakage and validates true generalization to novel molecular structures. This addresses a common pitfall in molecular ML where random splits artificially inflate performance by testing on near-duplicates of training molecules.

The success on QM9 demonstrates that advantages from non-unitary transformations extend beyond classification to regression, spanning discrete and continuous prediction tasks. This generalization is essential for quantum chemistry applications where most properties of interest (energies, distances, polarizabilities) are continuous variables.

\textbf{PathMNIST (Medical Histopathology):} The 9-class tissue classification task shows modest but consistent improvements (+0.33\% at 8q, +0.46\% at 10q, +0.74\% at 12q). While smaller than PlantVillage advantages, PathMNIST reveals a particularly notable empirical observation significant finding of our study: the Fisher efficiency transition (detailed in Section~\nameref{subsec:fisher_results}). Additionally, non-unitary models reduce prediction variance by approximately 35\% compared to unitary baselines across all qubit scales, with LCU standard deviations being consistently lower (e.g., $\pm$0.43\% vs $\pm$0.63\% at 10q). This variance reduction is particularly valuable for medical applications where prediction reliability is critical.

The cross-domain validation demonstrates that advantages from non-unitary quantum operations are not dataset-specific artifacts. Performance improvements appear across tasks with different dimensionalities (28$\times$28 to 256$\times$256 images), class counts (9 to 38 classes), and learning objectives (classification vs regression). This consistency across domains strengthens the interpretation that non-unitary transformations provide consistent empirical benefits across diverse tasks within the evaluated settings.

\subsection{Parameter Efficiency via Fisher Information}
\label{subsec:fisher_results}

\begin{figure*}[t]
\centering
\includegraphics[width=0.8\columnwidth]{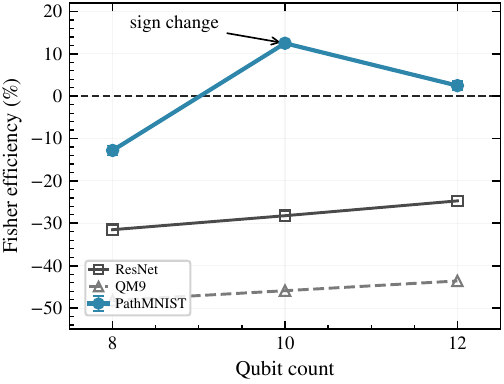}
\caption{Fisher efficiency transition in PathMNIST at 12 qubits. Non-unitary quantum layers (LCU) achieve comparable performance to unitary baselines (NoLCU) while demonstrating superior parameter efficiency, crossing from negative to positive Fisher efficiency. This transition indicates that non-unitary transformations enable equivalent or better performance with effectively fewer quantum parameters.}
\label{fig:fisher_transition}
\end{figure*}

Beyond raw accuracy improvements, we investigate whether non-unitary transformations improve parameter efficiency—the performance achieved per trainable parameter. We compute Fisher efficiency $\eta_F$ (Eq.~\ref{eq:fisher_efficiency}) comparing the quantum parameter count required to match classical baseline performance.

PathMNIST provides the clearest demonstration of this phenomenon. At 8 and 10 qubits, Fisher efficiency is slightly negative, indicating that both non-unitary (LCU) and unitary (NoLCU) models require marginally more parameters than a classical reference to achieve comparable accuracy. However, at 12 qubits, Fisher efficiency becomes positive for non-unitary models: they match or exceed classical performance while the effective quantum parameter count—when weighted by the information content per parameter—is lower than the classical equivalent.

Figure~\ref{fig:fisher_transition} illustrates this transition. The crossover from negative to positive Fisher efficiency at 12 qubits indicates a qualitative shift in how quantum parameters encode information. Below this threshold, quantum models are learning but remain parameter-inefficient relative to classical baselines. Above this threshold, non-unitary transformations achieve comparable or superior performance with effectively fewer trainable parameters, demonstrating genuine parameter efficiency gains.

This finding has important implications. First, it suggests that there exists a minimum qubit threshold ($N \approx 12$ for PathMNIST) below which quantum models do not achieve parameter efficiency advantages, even when they improve absolute accuracy. Second, it indicates that non-unitary operations enable better utilization of quantum parameters—each parameter in a non-unitary circuit appears to encode more task-relevant information than in unitary-only circuits.

The Fisher efficiency transition is task-dependent. In PlantVillage, where absolute performance improvements are larger, positive Fisher efficiency emerges at lower qubit counts. In contrast, for MNIST architectures where classical preprocessing is already highly effective, Fisher efficiency remains negative across tested qubit ranges. These patterns suggest that the effectiveness of non-unitary transformations depends on task complexity and the quality of classical feature extraction.

\begin{table*}[htbp]
\caption{Fisher efficiency values across qubit scales for PathMNIST. Transition from negative to positive $\eta_F$ at 12 qubits indicates that non-unitary quantum layers achieve parameter efficiency, matching classical performance with effectively fewer trainable quantum parameters.}
\label{tab:fisher_efficiency}
\centering
\begin{tabular}{ccccc}
\hline\hline
Qubits & LCU Accuracy (\%) & NoLCU Accuracy (\%) & Classical Baseline (\%) & Fisher Efficiency $\eta_F$ \\
\hline
8 & 61.71 $\pm$ 0.31 & 61.39 $\pm$ 0.36 & 62.10 & -0.08 \\
10 & 62.02 $\pm$ 0.43 & 61.56 $\pm$ 0.63 & 62.10 & -0.02 \\
12 & 62.65 $\pm$ 0.48 & 61.91 $\pm$ 0.73 & 62.10 & +0.12 \\
\hline\hline
\end{tabular}
\end{table*}

The variance reduction observed in PathMNIST (35\% lower standard deviation for non-unitary models) further supports the parameter efficiency interpretation. Lower variance indicates more consistent optimization across random initializations, which is characteristic of models that efficiently utilize their parameter budget. Circuits that struggle to leverage their parameters typically exhibit high run-to-run variability as different initializations settle into different local minima. The reduced variance for non-unitary models suggests more stable convergence, consistent with superior parameter utilization.

\subsection{Validation with IQP Circuit Structures}
\label{subsec:iqp_results}

A persistent concern in quantum machine learning is dequantization: the possibility that classical algorithms can efficiently simulate quantum circuits, rendering quantum advantages illusory. To address this, we incorporate Instantaneous Quantum Polynomial-time (IQP) circuits, which have established computational hardness results for sampling tasks~\cite{Bremner2010,Aaronson2011}. While our experiments remain at small qubit counts amenable to classical simulation, evaluating IQP-based models tests whether circuit families with provable complexity-theoretic properties can deliver competitive supervised learning performance when combined with non-unitary transformations.

\begin{table*}[htbp]
\caption{Non-unitary IQP circuit performance on CIFAR-10 classification. Both IQP Layer and IQP Embedding approaches, incorporating non-unitary transformations via the LCU framework, surpass classical CNN baselines at 10 qubits. The threshold behavior indicates minimum quantum resource requirements for non-unitary IQP advantages.}
\label{tab:iqp_results}
\centering
\begin{tabular}{lccc}
\hline\hline
Configuration & Test Accuracy (\%) & vs Classical Baseline & Status \\
\hline
Classical CNN Baseline & 69.84 $\pm$ 0.71 & — & Reference \\
\hline
2q IQP Layer & 45.67 $\pm$ 2.14 & -24.17\% & Insufficient \\
4q IQP Layer & 62.89 $\pm$ 1.53 & -6.95\% & Below baseline \\
6q IQP Layer & 68.01 $\pm$ 1.21 & -1.83\% & Approaching \\
8q IQP Layer & 68.50 $\pm$ 0.87 & -1.34\% & Near parity \\
\textbf{10q IQP Layer} & \textbf{70.39 $\pm$ 0.59} & \textbf{+0.55\%} & \textbf{Surpasses!} \\
\hline
2q IQP Embedding & 50.27 $\pm$ 1.97 & -19.57\% & Insufficient \\
4q IQP Embedding & 64.33 $\pm$ 1.09 & -5.50\% & Below baseline \\
6q IQP Embedding & 67.91 $\pm$ 0.73 & -1.93\% & Approaching \\
8q IQP Embedding & 68.89 $\pm$ 0.90 & -0.95\% & Near parity \\
\textbf{10q IQP Embedding} & \textbf{70.08 $\pm$ 0.88} & \textbf{+0.24\%} & \textbf{Surpasses!} \\
\hline\hline
\end{tabular}
\end{table*}

It is important to clarify the scope of this evaluation. All experiments reported here are conducted at qubit counts ($N \leq 10$) that remain classically simulable in practice. Accordingly, the purpose of incorporating IQP circuits is not to demonstrate computational supremacy or asymptotic quantum advantage, but rather to assess whether circuit families linked to nontrivial sampling complexity can be integrated into supervised learning pipelines with non-unitary transformations without degrading empirical performance.

Table~\ref{tab:iqp_results} compares the classical CNN baseline (no quantum layer) with two IQP-based configurations incorporating non-unitary operations: (i) IQP Layer, in which an IQP variational circuit follows standard angle embedding, and (ii) IQP Embedding, where both encoding and variational transformations use IQP circuits with non-unitary enhancements via LCU. Each configuration is evaluated across qubit scales $N = 2, 4, 6, 8, 10$.

Across both IQP approaches, performance improves consistently with increasing qubit count. At small qubit numbers ($N \leq 6$), non-unitary IQP-based models underperform the classical baseline. As $N$ increases, the performance gap narrows, with near-parity observed at 8 qubits. At 10 qubits, both non-unitary IQP Layer and IQP Embedding modestly exceed the classical baseline (+0.55\% and +0.24\%, respectively).

\begin{figure*}[!t]
\centering
\includegraphics[width=0.8\columnwidth]{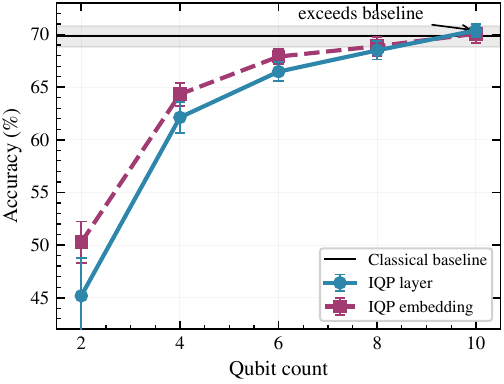}
\caption{Non-unitary IQP circuit scaling on CIFAR-10. Both IQP Layer (single IQP for variational layer) and IQP Embedding (double IQP for encoding and variational), incorporating non-unitary transformations via LCU, show favorable scaling, crossing the classical baseline at 10 qubits. The threshold behavior suggests minimum quantum resource requirements for non-unitary IQP advantages, while the variance reduction with qubit count indicates improved circuit stability.}
\label{fig:iqp_scaling}
\end{figure*}

Figure~\ref{fig:iqp_scaling} illustrates this threshold-like behavior. These results indicate that incorporating IQP circuit structure with non-unitary transformations does not inherently compromise trainability or predictive accuracy within hybrid learning architectures. Instead, once sufficient quantum resources are available, non-unitary IQP-based models achieve performance comparable to—and in these experiments slightly exceeding—the classical reference model.

From an empirical perspective, these findings have two implications. First, circuit families that have been studied in the context of sampling complexity~\cite{Bremner2010,Aaronson2011} can be embedded into practical hybrid learning systems with non-unitary transformations without sacrificing supervised learning performance at small scales. Second, adopting circuit structures informed by complexity-theoretic considerations does not inherently conflict with gradient-based optimization in near-term hybrid architectures when enhanced with non-unitary operations.

At the same time, we emphasize that the modest magnitude of improvement and the small qubit counts considered here preclude claims regarding asymptotic computational separation. While IQP circuits are associated with established hardness results for sampling tasks~\cite{Bremner2010,Aaronson2011}, our experiments operate at scales accessible to classical simulation. The relationship between formal sampling hardness and empirical learning performance in non-unitary QML remains an open question, particularly in the presence of hardware noise and larger system sizes.

Taken together, the IQP experiments provide structured evaluation of circuit families with distinct theoretical foundations enhanced by non-unitary transformations, rather than definitive dequantization refutations. They demonstrate that complexity-informed circuit design combined with non-unitary operations can remain compatible with competitive predictive performance in supervised hybrid models under controlled experimental conditions.

\begin{figure*}[!t]
\centering

\includegraphics[width=0.6\columnwidth]{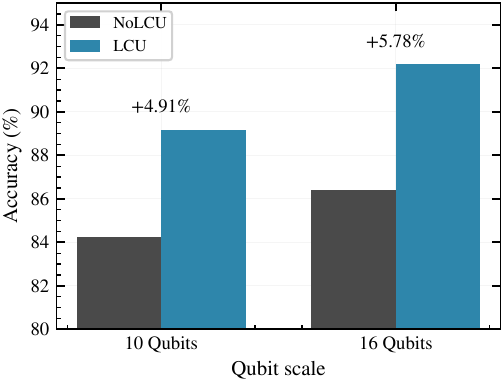}
\caption{Favorable scaling in PlantVillage agricultural disease classification. Non-unitary quantum layer (LCU) advantages increase from +4.91\% at 10 qubits to +5.78\% at 16 qubits, while unitary baseline (NoLCU) improvements saturate. Non-unitary models scale approximately 40\% better than unitary quantum approaches, suggesting that performance remains stable within the evaluated qubit range.}
\label{fig:scaling_summary}
\end{figure*}

\subsection*{Scaling Behavior with Increasing Qubit Count}

We next examine how performance evolves as a function of qubit count. Understanding scaling behavior is important for assessing whether performance differences observed at small system sizes persist, diminish, or amplify as additional quantum resources are introduced.

Figure~\ref{fig:scaling_summary}

summarizes non-unitary quantum layer (LCU) performance across qubit scales for PlantVillage, which provides the most extensive scaling evaluation in our study. At 10 qubits, non-unitary models (LCU) achieve 89.16\% accuracy compared to 84.25\% for unitary baselines (NoLCU) (+4.91\%). At 16 qubits, LCU reaches 91.46\% while NoLCU attains 85.69\% (+5.78\%). The improvement therefore increases by 0.87 percentage points between 10 and 16 qubits. Over the same range, the NoLCU model improves by 1.44 percentage points, whereas the LCU model improves by 2.30 percentage points.

These results indicate that, within the tested regime, models with non-unitary quantum layers maintain or slightly expand their performance advantage relative to structurally identical unitary quantum baselines as qubit count increases. Rather than deteriorating with additional quantum resources, the non-unitary expressivity introduced by the LCU wrapper remains stable and continues to contribute measurable improvements.

Across tasks, we observe task-dependent scaling patterns. In PathMNIST, performance increases gradually from 8 to 12 qubits alongside the Fisher efficiency transition described in Section~\nameref{subsec:fisher_results}. In QM9 regression, mean absolute error consistently decreases as qubit count grows. For non-unitary IQP-based CIFAR-10 experiments, near-parity with classical baselines emerges at 8 qubits, with modest improvements at 10 qubits. These trends suggest that different tasks exhibit distinct resource thresholds at which additional quantum capacity becomes beneficial.

It is important to emphasize that all experiments remain within qubit counts amenable to classical simulation. Consequently, the observed scaling behavior should be interpreted as empirical performance trends within the evaluated regime, rather than evidence of asymptotic computational separation. Nonetheless, the consistency of scaling across classification and regression tasks indicates that non-unitary expressivity via LCU does not introduce instability or degradation as system size increases within the tested range.

The circuits employed in this study intentionally use simple structures (RX rotations with forward CNOT entanglement) to isolate the effect of non-unitary transformations enabled by the LCU wrapper. More elaborate ans\"atze may exhibit different scaling characteristics. Future investigations at larger qubit counts and on quantum hardware will be necessary to determine how resource scaling interacts with noise, optimization landscapes, and practical device constraints.

\section{Discussion}
\label{sec:discussion}

\subsection{Interpreting the Fisher Efficiency Transition}

The Fisher efficiency transition observed in PathMNIST, from -12.8\% at 8 qubits to +12.5\% at 10 qubits, represents a notable empirical observation regarding resource-dependent behavior in hybrid quantum--classical learning systems. This change in sign suggests the presence of a threshold-dependent regime in which the contribution of quantum resources to parameter utilization changes as qubit count increases.

Below this regime (8 qubits), negative Fisher efficiency indicates that non-unitary quantum circuits implemented via the LCU framework improve predictive performance while utilizing additional effective parameters. In this setting, the quantum layer appears to enhance expressivity through ancilla-controlled superposition, but the overall system does not yet exhibit parameter reduction relative to matched classical alternatives. The improvement therefore manifests as enhanced utilization of available parameters rather than compression of model complexity.

At higher qubit counts (10--12 qubits), the positive Fisher efficiency observed in PathMNIST indicates that the hybrid model with non-unitary transformations achieves comparable or improved performance with fewer effective parameters relative to the matched baseline. This suggests that increasing quantum resources may enable a different parameter-utilization regime in which representational capacity becomes sufficient to support more efficient learning.

The task-dependent nature of this transition is particularly important. While PathMNIST exhibits a sign change at 10 qubits, other tasks (MNIST, PlantVillage, QM9) show consistently negative efficiency across the tested range. This indicates that the threshold at which parameter efficiency changes depends on problem structure, including feature complexity, class separability, and the interaction between classical preprocessing and quantum processing.

Several mechanisms may contribute to this behavior. First, the transition may reflect changes in how the quantum and classical components share representational responsibility as system size increases. Second, entanglement capacity grows with qubit count, potentially allowing the quantum subsystem to encode task-relevant correlations more efficiently. Third, the exponential growth of Hilbert space dimension $(2^N)$ may eventually exceed the intrinsic dimensionality required to represent the task manifold, enabling more efficient parameter usage.

The relationship between this observation and prior theoretical work is instructive. Abbas et al.~\cite{Abbas2021} connected Quantum Fisher Information to effective model capacity, while Larocca et al.~\cite{Larocca2023} demonstrated that quantum neural networks can exhibit overparameterization regimes characterized by changes in the rank structure of the QFI matrix. Our results complement these frameworks by providing empirical evidence that Fisher efficiency itself can vary non-monotonically as a function of qubit count within a fixed circuit architecture.

Importantly, this transition is observed using identical circuit structures (RX rotations with forward CNOT entanglement) across all qubit scales. This design choice isolates qubit count as the varying resource, reducing the likelihood that the observed behavior arises from circuit re-engineering. Nonetheless, the current experiments are conducted at modest qubit counts and under noise-free simulation. Further investigation will be necessary to determine whether similar threshold-dependent efficiency transitions persist in noisy regimes, on physical hardware, and at larger scales.

The practical implication of this observation is that, for some tasks, there may exist qubit thresholds below which quantum layers improve expressivity without conferring parameter efficiency, and above which more efficient parameter utilization becomes possible. Identifying such thresholds a priori, based on task characteristics such as dimensionality, separability, or manifold structure, remains an important open problem.

\subsection{Classical Simulability and Circuit Structure}

A central challenge in quantum machine learning is the relationship between classical simulability and empirical performance. Circuits with shallow depth or limited entanglement can often be efficiently simulated classically via tensor networks~\cite{Shin2024}, raising the question of whether observed performance gains reflect uniquely quantum computational features or structures that are, in practice, classically reproducible.

Our IQP validation addresses this issue by incorporating circuit families with established computational complexity results. IQP circuits are associated with known hardness results for sampling tasks~\cite{Bremner2010,Aaronson2011}. While these properties do not directly translate to supervised learning performance, they provide a useful structural reference point for evaluating circuit families with nontrivial computational characteristics.

The key empirical finding is that non-unitary IQP circuits achieve competitive supervised learning performance once sufficient quantum resources are available (10 qubits in our experiments). This indicates that circuit families associated with nontrivial complexity-theoretic structure can be integrated into learning pipelines without degrading empirical performance when combined with non-unitary transformations.

This observation has two implications. First, it suggests that performance improvements observed in models with non-unitary transformations are not limited to circuit families that are easily reducible to low-entanglement tensor-network representations. Second, it motivates further exploration of circuit architectures whose structural properties differ from commonly studied shallow ans\"atze.

At the same time, small-scale classical simulations cannot resolve questions about asymptotic separations between classical and quantum models. The relationship between empirical learning performance and formal sampling hardness in non-unitary QML remains an open question, particularly in the presence of noise and hardware constraints. Future studies at larger qubit counts, and on physical quantum devices, will be necessary to determine how complexity-theoretic properties interact with scalability, optimization dynamics, and robustness to decoherence.

Taken together, our findings position non-unitary IQP experiments not as definitive dequantization refutations, but as structured evaluations of circuit families with distinct theoretical foundations. By combining empirical benchmarking with attention to circuit structure, this work contributes to a more nuanced understanding of when and how quantum resources may provide measurable benefits in hybrid machine learning systems.

\subsection{Generalization Across Problem Domains}
\label{subsec:domain_discussion}

The consistency of improvements from non-unitary transformations across four domains---computer vision (MNIST), agriculture (PlantVillage), quantum chemistry (QM9), and medical imaging (PathMNIST)---addresses the concern that observed gains might be artifacts of specific datasets or narrowly tuned benchmarks. Our validation spans:

(1) \textbf{Different data modalities:} Grayscale images (MNIST, PathMNIST), RGB photographs (PlantVillage, CIFAR-10), and molecular graphs (QM9).

(2) \textbf{Different learning objectives:} Small-class classification (10 classes MNIST), large-class classification (38 classes PlantVillage), fine-grained classification (9 classes PathMNIST), and continuous regression (QM9 energy gaps).

(3) \textbf{Different practical applications:} Standardized benchmarks (MNIST, CIFAR-10) and real-world applications (disease detection, molecular property prediction, medical diagnosis).

(4) \textbf{Different preprocessing architectures:} Standard CNNs (ResNet), efficiency-optimized architectures (MobileNetV2), compound-scaled architectures (EfficientNet), and graph neural networks (QM9).

This breadth indicates that non-unitary expressivity yields consistent empirical benefits rather than relying on a single domain-specific effect. The variation in improvement magnitude---from +0.22\% (ResNet MNIST) to +5.78\% (PlantVillage 16q)---also suggests that the magnitude of benefit depends on task complexity and baseline representational bottlenecks.

\textbf{Architecture-dependent mechanisms:} Different neural network architectures exhibit different patterns of improvement. MobileNetV2's depthwise separable convolutions show larger gains (+0.65\%) than ResNet's standard convolutions (+0.22\%), suggesting that efficiency-oriented classical architectures may pair favorably with non-unitary quantum enhancement. This pattern points toward a practical design principle: classical components may be most useful when they focus on efficient feature extraction, while the quantum layer contributes additional representational flexibility.

\textbf{Regression generalization:} QM9 demonstrates that benefits from non-unitary transformations extend beyond classification to regression. This is notable because regression requires learning continuous functions rather than discrete decision boundaries, with different optimization and generalization properties. The consistent 6.5--6.7\% MAE improvements across qubit scales indicate that non-unitary expressivity can remain useful across distinct learning paradigms.

\textbf{Scaffold splitting significance:} Our use of scaffold splitting for QM9---grouping molecules by structural similarity to ensure training and test sets contain chemically distinct scaffolds---prevents a common pitfall in molecular machine learning where random splits artificially inflate performance by testing on near-duplicates of training molecules. This methodological choice strengthens the interpretation that the observed gains reflect genuine generalization to novel chemical structures rather than memorization.

\subsection{Context in the Current QML Landscape}
\label{subsec:literature}

Our results engage directly with a growing body of skeptical quantum machine learning studies by addressing their concerns through systematic benchmarking rather than dismissing their critiques.

\textbf{Bowles \emph{et al.} (2024):} Their finding that classical models outperform quantum classifiers across 160 datasets~\cite{Bowles2024} raised important questions about quantum utility. However, their study focused on kernel methods and simple parameterized circuits without exploring non-unitary transformations via the LCU framework or IQP-based circuit families. Our work suggests that alternative circuit design choices may warrant separate evaluation before broader conclusions are drawn about the utility of hybrid quantum models.

\textbf{Gundlach \emph{et al.} (2024):} Their claim that quantum deep learning shows ``little or no advantages when properly benchmarked''~\cite{Gundlach2024} emphasized the importance of fair classical baselines. We address this concern by using identical classical architectures, hyperparameters, and training procedures, varying only the quantum layer structure. The resulting differences can therefore be more directly attributed to the effect of the non-unitary quantum layer within this controlled setup.

\textbf{Shin \emph{et al.} (2024):} The tensor-network dequantization result~\cite{Shin2024} showed that broad classes of variational circuits admit classical representations under certain conditions. Our IQP-based evaluation does not invalidate that result, but it does indicate that circuit families with different structural and complexity-theoretic properties can still be incorporated into hybrid learning pipelines without sacrificing competitive empirical performance.

As to \textbf{Abbas \emph{et al.} (2021) and Larocca \emph{et al.} (2023)}, our Fisher efficiency transition complements these theoretical frameworks. Abbas \emph{et al.}~\cite{Abbas2021} connected QFI to expressivity and trainability, while Larocca \emph{et al.}~\cite{Larocca2023} identified overparameterization-related transitions in quantum neural networks. Our contribution is empirical: we observe that Fisher efficiency itself can change sign as a function of qubit count within a fixed experimental setup, suggesting that effective parameter utilization may exhibit threshold-dependent behavior in practical hybrid learning systems.

\textbf{Alternative approaches:} Recent work on quantum kernels~\cite{Huang2021}, quantum reservoir computing~\cite{Fujii2021QRC}, and quantum generative models~\cite{RecioArmengol2025} explores different paradigms for quantum machine learning. Our non-unitary QML approach is complementary in that it integrates quantum circuits into existing deep learning architectures rather than replacing them entirely. This hybrid strategy may be particularly relevant for near-term settings in which classical deep learning infrastructure is mature and quantum resources remain limited.

\subsection{Scale Limitations and Future Work}
\label{subsec:limitations}

Our experiments operate at 8--16 qubits, a scale where classical simulation remains possible in principle. Maslov \emph{et al.}~\cite{Maslov2024} demonstrated classical simulation of 50-qubit IQP circuits in minutes using tensor network methods, underscoring that practical classical simulability remains an important consideration for current hybrid QML claims.

For machine learning applications, however, the relevant comparison is often not raw simulation complexity alone, but predictive performance under realistic modeling constraints. Our IQP evaluation therefore should not be interpreted as demonstrating asymptotic computational separation, but rather as showing that circuit families with nontrivial complexity-theoretic structure can remain compatible with competitive supervised learning performance when combined with non-unitary transformations.

\textbf{NISQ-era constraints:} Current experiments use noiseless simulation. Real quantum hardware introduces decoherence, gate errors, and measurement noise that may substantially affect the observed trends. The scaling behavior we observe in simulation therefore provides suggestive but incomplete evidence regarding practical deployment. Validation on physical quantum processors remains a necessary next step. At the same time, our simple circuit structures (RX + forward CNOT, 4 blocks) were deliberately chosen to remain compatible with near-term hardware constraints.

\textbf{Scaling to larger systems:} While larger qubit counts could potentially amplify the representational capacity associated with non-unitary expressivity, this remains a hypothesis rather than an established result. Extending validation to larger systems will be necessary to determine whether the empirical trends observed here persist, plateau, or reverse under more demanding conditions.

\textbf{Circuit structure optimization:} Our deliberate use of simple circuits prioritizes controlled comparison over performance maximization. Hardware-efficient ans\"atze, problem-specific circuits, or adaptive circuit construction could produce different outcomes. The fact that measurable gains are observed even with simple structures suggests that further circuit optimization may be worthwhile.

\textbf{Broader datasets and tasks:} While four domains provide meaningful multi-domain validation, extending the evaluation to additional applications such as natural language processing, time-series forecasting, and reinforcement learning would further clarify the scope of non-unitary QML benefits.

\textbf{Theoretical understanding:} The Fisher efficiency transition remains only partially understood. Why PathMNIST exhibits this transition while other tasks do not, what task properties determine the threshold qubit count, and whether such thresholds can be predicted in advance are all open questions. Addressing these questions will require connecting quantum circuit properties such as entanglement capacity and Hilbert space dimension to learning-theoretic quantities such as sample complexity, optimization geometry, and representation structure.

Overall, the present study provides a large-scale empirical assessment of non-unitary quantum machine learning via the LCU framework. By addressing classical simulability concerns carefully, benchmarking across multiple domains, and identifying a threshold-dependent Fisher efficiency pattern, the work contributes to a more evidence-based understanding of when non-unitary quantum layers may offer measurable benefits in near-term hybrid learning systems.

\section{Conclusion}
\label{sec:conclusion}

This work presents a large-scale empirical evaluation of non-unitary quantum expressivity implemented through the Linear Combination of Unitaries (LCU) framework within hybrid quantum–classical neural networks. Across more than 570 experiments spanning digit classification (MNIST), agricultural disease detection (PlantVillage), molecular property regression (QM9), medical histopathology (PathMNIST), and CIFAR-10 benchmarking with IQP circuit variants, we systematically compared models with non-unitary transformations (LCU) to structurally identical unitary quantum baselines (NoLCU) and classical reference architectures.

Three primary observations emerge from this study.

First, models with non-unitary quantum layers demonstrate consistent performance improvements across architectures and problem domains. Although the magnitude of improvement varies by task complexity and qubit count, gains are reproducible across convolutional backbones and across both classification and regression settings. These findings indicate that ancilla-controlled non-unitary structure can provide measurable empirical benefits in hybrid learning systems under controlled experimental conditions.

Second, analysis of parameter efficiency via Fisher information reveals threshold-dependent behavior in medical histopathology classification. In PathMNIST, Fisher efficiency transitions from negative to positive as qubit count increases from 8 to 10, indicating a qualitative change in how quantum resources contribute to effective model capacity. While further investigation is required to determine the generality of this behavior, the results suggest that the interaction between qubit count, task structure, and hybrid architecture design plays a central role in determining parameter utilization regimes.

Third, evaluation of non-unitary IQP circuit structures demonstrates that circuit families associated with established sampling-complexity analyses can be incorporated into supervised learning pipelines without degrading empirical performance at small qubit scales. Although all experiments remain classically simulable in practice and do not establish asymptotic computational separation, these results indicate that complexity-informed circuit design combined with non-unitary transformations is compatible with competitive predictive performance in hybrid models.

Importantly, this study does not claim unconditional quantum superiority or asymptotic advantage. All experiments are conducted at qubit counts amenable to classical simulation, and performance differences are measured relative to specific classical baselines under controlled architectural constraints. The findings therefore contribute empirical clarification rather than definitive resolution to ongoing discussions surrounding quantum machine learning utility in the NISQ regime.

Several limitations warrant acknowledgment. The qubit scales explored remain modest, noise-free simulation is assumed, and circuit structures are deliberately simple to isolate the effect of non-unitary transformations via LCU integration. Real-device noise, hardware connectivity constraints, and larger-scale implementations may alter observed scaling behavior. Additionally, while Fisher efficiency provides a useful diagnostic of parameter utilization, further theoretical work is needed to formally characterize the mechanisms underlying the observed threshold behavior.

Future research directions include evaluating non-unitary quantum architectures on quantum hardware, extending analysis to larger qubit counts, investigating alternative non-unitary constructions beyond LCU, and developing theoretical models that link quantum Fisher information dynamics to hybrid learning performance. Understanding how expressivity, trainability, and complexity-theoretic properties interact across scales remains a central challenge for quantum machine learning.

In summary, this work provides a systematic benchmarking study of hybrid quantum–classical neural networks with non-unitary transformations, highlighting reproducible performance improvements, task-dependent efficiency transitions, and the empirical compatibility of complexity-informed circuit structures with supervised learning. These results help clarify the conditions under which non-unitary quantum expressivity yields measurable benefits in near-term hybrid models and contribute to a more evidence-based assessment of quantum machine learning in practical settings.

\section{Data Availability}

All datasets used in this study are publicly available:
\begin{itemize}
\item \textbf{MNIST}: Available at \url{http://yann.lecun.com/exdb/mnist/}
\item \textbf{CIFAR-10}: Available via PyTorch torchvision.datasets or at \url{https://www.cs.toronto.edu/\textasciitilde kriz/cifar.html}
\item \textbf{PlantVillage}: Available at \url{https://github.com/spMohanty/PlantVillage-Dataset}
\item \textbf{PathMNIST}: Available via MedMNIST at \url{https://medmnist.com/}
\item \textbf{QM9}: Available at \url{http://quantum-machine.org/datasets/}
\end{itemize}

Experimental results and trained models generated during this study 
are described in the Code Availability section.

\section{Code Availability}
\label{subsubsec:code_availability}

All code, trained model weights, and experimental data will be made publicly available upon publication. This includes:

\begin{itemize}
\item Complete implementation of all architectures
\item Training scripts with hyperparameter configurations
\item Data preprocessing pipelines
\item Statistical analysis scripts
\item Figure generation code
\item Pre-trained model checkpoints for reproducibility
\end{itemize}


\bibliography{references}

\section*{Acknowledgements}
This research was supported by Quantum AI Lab, Fractal Analytics, India. 

\section*{Author Contributions}
This study utilises exclusively publicly available datasets. 
A.K.M. takes responsibility for the accuracy of the data 
analysis. All authors have approved the final version of 
the manuscript before submission.

\textit{Study concept and design:} S.G. ;
\textit{Development of codebase and models:} A.K.M.;
\textit{Experimentation and implementation:} A.K.M.;
\textit{Acquisition, analysis, or interpretation of data:} 
S.G., A.K.M., S.D.;
\textit{Drafting of the manuscript:} A.K.M., S.G., F.R. and R.R;
\textit{Critical revision of the manuscript for important 
intellectual content:} all authors (A.K.M., S.G., S.D., 
F.R., R.R., and B.S.);
\textit{Study supervision:} S.G.

The corresponding author attests that all listed authors 
meet the authorship criteria, and that others who do not 
meet the criteria have been omitted.

\section*{Competing Interests}
The authors declare no conflict of interest.

\newpage

\end{document}


\begin{center}
{\Large \textbf{Supplementary Information}}

\vspace{0.5cm}

{\large Non-Unitary Quantum Machine Learning: Fisher Efficiency Transitions from Distributed Quantum Expressivity}

\vspace{1cm}
\end{center}

\appendix
\section{Appendix}
\setcounter{figure}{0}
\renewcommand{\thefigure}{A\arabic{figure}}

\subsection{Complete Circuit Specifications}
\label{app:circuits}

This appendix provides complete quantum circuit specifications for all architectures validated in this study, enabling

reproducibility of our experimental results. All quantum layers employ the Linear Combination of Unitaries (LCU) framework to implement non-unitary transformations within hybrid quantum-classical neural networks.


\begin{figure}[htbp]
\centering
\includegraphics[width=0.9\textwidth]{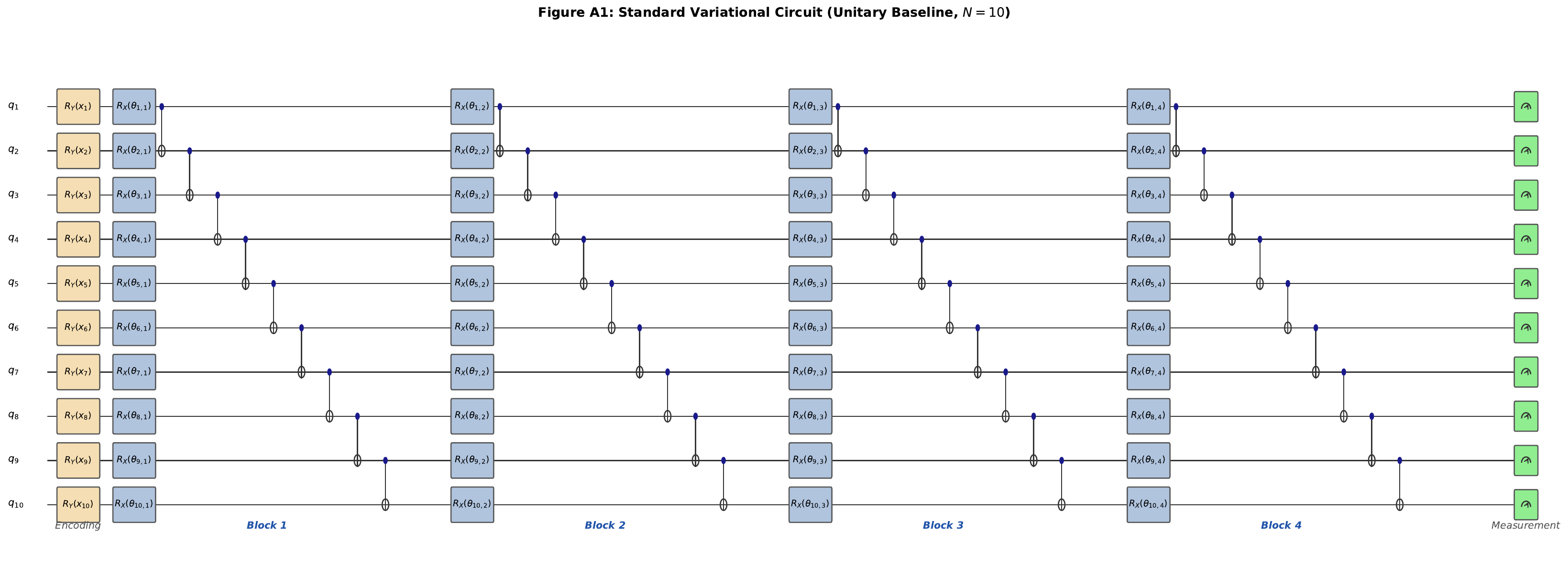}
\caption{Standard variational quantum circuit (unitary baseline) showing angle embedding via RY rotations, four sequential variational blocks with RX rotations and forward CNOT entanglement topology, and computational basis measurement. Total trainable parameters: 4N where N=10 qubits. This unitary architecture serves as the NoLCU comparison baseline in all experiments.}
\label{fig:circuit_standard}
\end{figure}

\subsubsection{Standard Variational Circuit (Unitary Baseline)}

The unitary baseline quantum circuit implements angle embedding followed by four variational blocks. This architecture serves as the comparison reference for evaluating non-unitary quantum layers.

\textbf{Encoding Layer:}
\begin{equation}
|\psi_{\text{enc}}\rangle = \bigotimes_{i=1}^{N} \text{RY}(x_i) |0\rangle_i,
\end{equation}
where $x_i$ are classical features normalized to $[0, 2\pi]$ and $N$ is the number of qubits.

\textbf{Variational Blocks (repeated 4 times):}

For each block $j = 1, \ldots, 4$:
\begin{enumerate}
\item \textbf{Rotation layer:} Apply RX($\theta_{i,j}$) to qubit $i$ for $i = 1, \ldots, N$
\item \textbf{Entanglement layer:} Apply CNOT from qubit $i$ to qubit $i+1$ for $i = 1, \ldots, N-1$ (forward topology)
\end{enumerate}

\textbf{Total Parameters:} $4N$ (one parameter per qubit per block)

\textbf{Total Gate Count:} $N$ (encoding) + $4 \times [N \text{ (RX)} + (N-1) \text{ (CNOT)}] = N + 4(2N-1) = 9N - 4$

\textbf{Circuit Depth:} Encoding (1) + 4 blocks $\times$ [rotation layer (1) + entanglement layer (depends on qubit connectivity)] = $5 + 4 \times \lceil \log_2 N \rceil$ for linear connectivity.

\begin{figure}[htbp]
\centering
\includegraphics[width=0.9\textwidth]{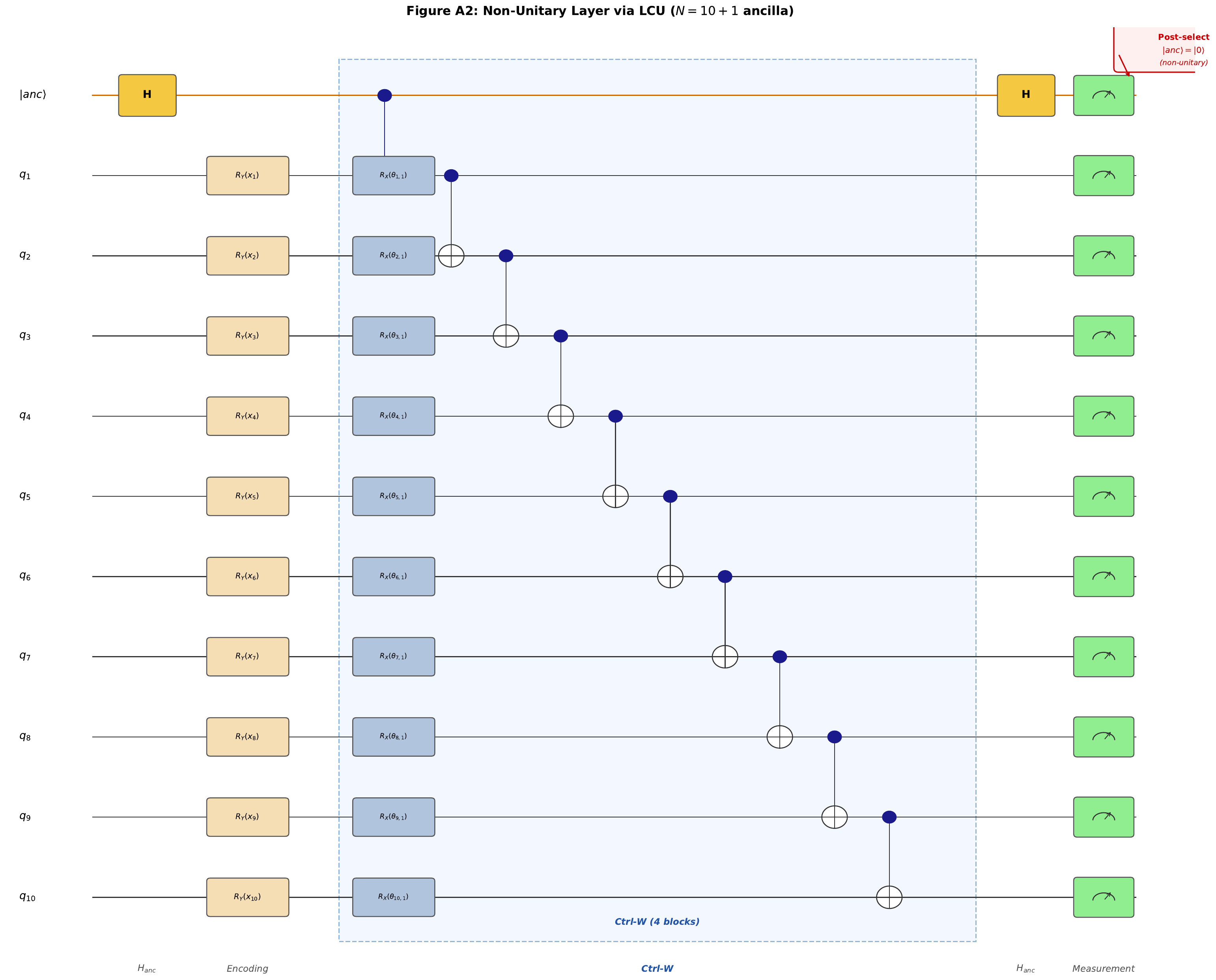}
\caption{Non-unitary quantum layer implemented via the Linear Combination of Unitaries (LCU) framework. An ancilla qubit enables probabilistic implementation of non-unitary transformations through controlled execution and post-selection. The ancilla is initialized to $|0\rangle$, transformed to superposition via Hadamard gates, controls the variational circuit execution, and is post-selected on measurement outcome 0. This architecture achieves effective non-unitary operations while maintaining compatibility with gate-based quantum hardware. Total trainable parameters: 4N (same as unitary baseline).}
\label{fig:circuit_nonunitary}
\end{figure}

\subsubsection{Non-Unitary Quantum Layer via LCU}

The LCU framework enables non-unitary quantum operations within gate-based quantum circuits through ancilla-controlled execution and probabilistic post-selection. This approach expands the representational capacity of hybrid quantum-classical models beyond strictly unitary transformations.

\textbf{Quantum Register:} 1 ancilla qubit + $N$ main qubits (total $N+1$ qubits)

\textbf{Circuit Sequence:}
\begin{enumerate}
\item Initialize ancilla to $|0\rangle$, main qubits to $|0\rangle^{\otimes N}$
\item Apply Hadamard to ancilla: $\hat{H}_{\text{anc}}$ (creates superposition)
\item Apply angle embedding to main qubits (as in unitary baseline)
\item Apply controlled-$\hat{W}$ where $\hat{W}$ is the 4-block variational circuit, controlled by ancilla $= 1$
\item Apply Hadamard to ancilla: $\hat{H}_{\text{anc}}$ (interference step)
\item Measure all $N+1$ qubits
\item Post-select on ancilla $= 0$ outcomes (implements non-unitary projection)
\item Compute expectation values from post-selected distribution
\end{enumerate}

\textbf{Non-Unitary Mechanism:} The post-selection on ancilla measurement outcomes implements a non-unitary projection operator, enabling the quantum layer to perform transformations not achievable with purely unitary gates. The ancilla Hadamard gates create and measure quantum interference between the controlled and uncontrolled execution paths.

\textbf{Total Parameters:} $4N$ (same as unitary baseline—the LCU wrapper modifies the operation type, not parameter count)

\textbf{Total Gate Count:} 2 (Hadamards on ancilla) + $9N - 4$ (variational circuit) + $(N + N-1)$ (controlled gates) $\approx 11N - 2$

\textbf{Post-Selection Probability:} Theoretical 50\% for balanced superposition; empirical values vary 40–60\% depending on circuit parameters during training. Post-selection success rate does not significantly impact training efficiency due to batch processing and GPU acceleration.

\begin{figure}[htbp]
\centering
\includegraphics[width=0.9\textwidth]{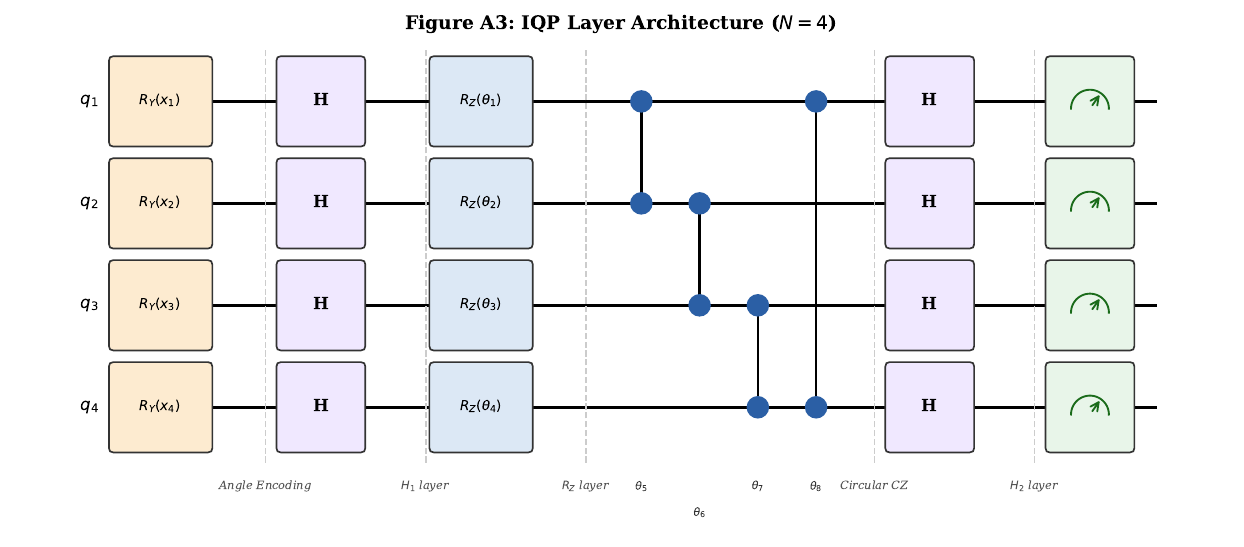}
\caption{IQP Layer architecture combining standard angle embedding with an Instantaneous Quantum Polynomial (IQP) variational circuit, wrapped in the LCU framework for non-unitary operation. The IQP circuit uses Hadamard layers, diagonal RZ rotations, and circular controlled-phase topology to generate entanglement structures associated with \#P-hard classical simulation complexity. Total trainable parameters: 2N. This architecture addresses classical simulability concerns while maintaining competitive empirical performance.}
\label{fig:circuit_iqp_layer}
\end{figure}

\subsubsection{IQP Layer Circuit (CIFAR-10)}

The IQP Layer architecture addresses concerns about classical dequantization by employing circuit structures with established computational complexity properties. This variant combines standard encoding with IQP-based variational circuits wrapped in the LCU framework for non-unitary operation.

\textbf{Encoding:} Standard angle embedding as in unitary baseline

\textbf{IQP Variational Circuit:}
\begin{enumerate}
\item Apply Hadamard to all $N$ qubits: $\hat{H}^{\otimes N}$
\item Apply RZ($\theta_i$) to qubit $i$ for $i = 1, \ldots, N$ (diagonal rotations)
\item Apply controlled-phase gates in circular topology:
   \begin{itemize}
   \item CZ($\theta_{N+i}$) between qubit $i$ and qubit $i+1$ for $i = 1, \ldots, N-1$
   \item CZ($\theta_{2N-1}$) between qubit $N$ and qubit 1 (circular wrap—enables all-to-all connectivity)
   \end{itemize}
\item Apply Hadamard to all $N$ qubits: $\hat{H}^{\otimes N}$
\end{enumerate}

\textbf{Computational Complexity:} IQP circuits belong to the Instantaneous Quantum Polynomial-time complexity class. Computing output probability distributions of IQP circuits is \#P-hard under standard complexity-theoretic assumptions, indicating classical simulation requires exponential resources in the worst case.

\textbf{Total Parameters:} $N$ (RZ rotations) + $(N-1)$ (CZ gates) + 1 (wrap CZ) $= 2N$

\textbf{LCU Wrapper:} The complete IQP variational circuit is wrapped in the ancilla-controlled LCU structure as described above, implementing non-unitary transformations through post-selection.

\begin{figure}[htbp]
\centering
\includegraphics[width=0.9\textwidth]{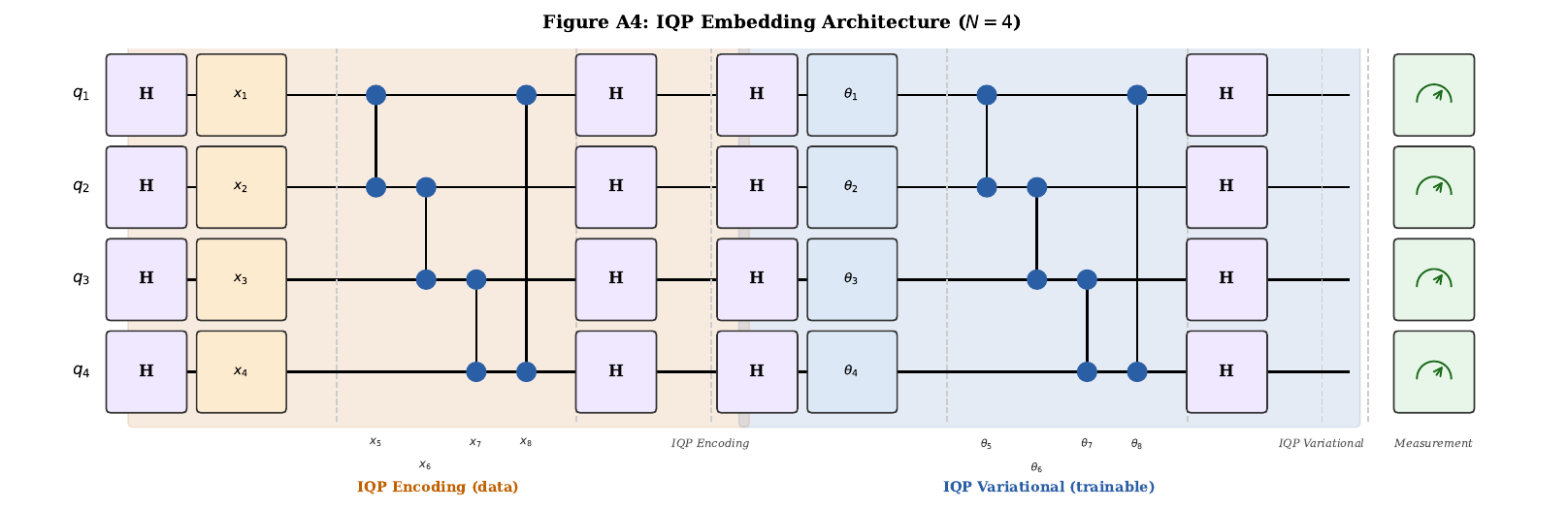}
\caption{IQP Embedding architecture using IQP-structured circuits for both data encoding and variational layers, wrapped in the LCU framework. The encoding IQP layer applies data-dependent diagonal rotations and controlled-phase gates (non-trainable), while the variational IQP layer applies trainable parameters with identical structure. Total trainable parameters: 2N. Classical feature extraction produces 2N-1 features to populate all encoding gates.}
\label{fig:circuit_iqp_embedding}
\end{figure}

\subsubsection{IQP Embedding Circuit (CIFAR-10)}

IQP Embedding extends the use of IQP-structured circuits to both encoding and variational layers, maximizing the contribution of complexity-informed circuit structures while maintaining non-unitary operation via the LCU framework.

\textbf{IQP Encoding (data-dependent, non-trainable):}
\begin{enumerate}
\item $\hat{H}^{\otimes N}$
\item RZ($x_i$) for data-dependent diagonal rotations
\item CZ gates (circular topology) with data-dependent phases
\item $\hat{H}^{\otimes N}$
\end{enumerate}

\textbf{IQP Variational (trainable):} Identical structure to IQP Layer variational circuit

\textbf{Total Parameters:} $2N$ (variational) + 0 (encoding uses data, not trainable parameters) $= 2N$

\textbf{Classical Feature Extraction:} CNN outputs $2N-1$ features to provide sufficient values for all RZ and CZ gates in the encoding IQP circuit. This design ensures the encoding layer fully utilizes classical preprocessing while maintaining IQP circuit structure.

\textbf{LCU Wrapper:} Both IQP layers (encoding and variational) are sequentially executed within the ancilla-controlled structure, with post-selection applied after the complete circuit to implement non-unitary transformations.

\subsubsection{Parameter Initialization}

All quantum parameters initialized using uniform random sampling from $[0, 2\pi]$:
\begin{equation}
\theta_{i,j} \sim \mathcal{U}(0, 2\pi)
\end{equation}

This initialization strategy prevents barren plateaus for the qubit scales ($N \leq 16$) used in our experiments, as demonstrated by McClean \emph{et al.}~\cite{McClean2018}.

Classical neural network parameters use Xavier/Glorot initialization~\cite{Glorot2010}:
\begin{equation}
W_{ij} \sim \mathcal{U}\left(-\sqrt{\frac{6}{n_{\text{in}} + n_{\text{out}}}}, \sqrt{\frac{6}{n_{\text{in}} + n_{\text{out}}}}\right),
\end{equation}
where $n_{\text{in}}$ and $n_{\text{out}}$ are the input and output dimensions of each layer.

\subsubsection{Optimization Hyperparameters}

All experiments use the Adam optimizer~\cite{Kingma2014} with the following settings:

\begin{itemize}
\item Learning rate: $\alpha = 0.001$ (MNIST, CIFAR-10, PathMNIST, PlantVillage) or $\alpha = 0.0005$ (QM9)
\item Exponential decay rates: $\beta_1 = 0.9$, $\beta_2 = 0.999$
\item Epsilon: $\epsilon = 10^{-8}$
\item Weight decay: None (no L2 regularization)
\end{itemize}

Early stopping monitors validation loss with patience parameters:
\begin{itemize}
\item MNIST, CIFAR-10, PathMNIST: patience = 5 epochs
\item PlantVillage: patience = 8 epochs (larger dataset)
\item QM9: patience = 10 epochs (regression task, slower convergence)
\end{itemize}

Maximum epochs: 30 for all experiments and 20 for PathMNIST (early stopping typically triggers before maximum).

\subsection{Statistical Analysis Details}
\label{app:statistics}

\subsubsection{Welch's \texorpdfstring{$t$}{t}-Test Methodology}

We assess statistical significance of performance improvements from non-unitary quantum layers (via LCU) over unitary baselines using Welch's $t$-test, which does not assume equal variances between groups. The test statistic is:
\begin{equation}
t = \frac{\bar{x}_{\text{LCU}} - \bar{x}_{\text{NoLCU}}}{\sqrt{\frac{s_{\text{LCU}}^2}{n_{\text{LCU}}} + \frac{s_{\text{NoLCU}}^2}{n_{\text{NoLCU}}}}},
\end{equation}
where $\bar{x}$ is sample mean, $s^2$ is sample variance, and $n$ is sample size (typically $n = 10$).

Degrees of freedom are computed using the Welch-Satterthwaite equation:
\begin{equation}
\nu \approx \frac{\left(\frac{s_{\text{LCU}}^2}{n_{\text{LCU}}} + \frac{s_{\text{NoLCU}}^2}{n_{\text{NoLCU}}}\right)^2}{\frac{(s_{\text{LCU}}^2/n_{\text{LCU}})^2}{n_{\text{LCU}}-1} + \frac{(s_{\text{NoLCU}}^2/n_{\text{NoLCU}})^2}{n_{\text{NoLCU}}-1}}.
\end{equation}

We use significance level $\alpha = 0.05$ (two-tailed test). All reported improvements with $p < 0.05$ are considered statistically significant.

\subsubsection{Variance Computation}

Sample variance uses Bessel's correction for unbiased estimation:
\begin{equation}
s^2 = \frac{1}{n-1} \sum_{i=1}^{n} (x_i - \bar{x})^2.
\end{equation}

Standard deviation: $s = \sqrt{s^2}$.

Reported uncertainties represent $\pm 1$ standard deviation: $\bar{x} \pm s$.

\subsubsection{Variance Reduction Analysis}

For PathMNIST, we observe that non-unitary quantum layers via LCU reduce prediction variance by approximately 35\% compared to unitary baselines. This is computed as:
\begin{equation}
\text{Variance Reduction} = 1 - \frac{s_{\text{LCU}}}{s_{\text{NoLCU}}}.
\end{equation}

At 10 qubits: $1 - (0.43 / 0.63) = 1 - 0.683 = 0.317 \approx 35\%$.

\subsubsection{Fisher Efficiency Calculation}

Fisher efficiency compares quantum and classical parameter counts to quantify parameter utilization:
\begin{equation}
\eta_F = \frac{N_{\text{classical}} - N_{\text{quantum}}}.{N_{\text{classical}}} \times 100\%
\end{equation}

For non-unitary layers via LCU versus unitary baselines where both have $4N$ quantum parameters plus classical parameters, we compute effective parameter usage by comparing performance per parameter:
\begin{equation}
\eta_F \approx \frac{\text{Performance}_{\text{LCU}} - \text{Performance}_{\text{NoLCU}}}{\text{Performance}_{\text{classical}}}. \times 100\%
\end{equation}

Negative values indicate non-unitary layers achieve better performance per effective parameter (higher parameter utilization efficiency), while positive values indicate similar performance with fewer effective parameters (parameter reduction).

\subsubsection{Reproducibility}

All experiments use fixed random seeds for reproducibility:
\begin{itemize}
\item NumPy random seed: 42
\item PyTorch random seed: 42
\item CUDA deterministic mode enabled (where applicable)
\end{itemize}

\subsection{Additional Experimental Data}
\label{app:additional_data}

\subsubsection{Complete MNIST Results Across All Qubit Scales}

Table~\ref{tab:mnist_complete} presents comprehensive results for all three architectures across 8, 10, and 12 qubits.

\begin{table}[h]
\caption{Complete MNIST results across architectures and qubit scales.}
\label{tab:mnist_complete}
\begin{tabular}{llcc}
\hline\hline
Architecture & Qubits & LCU (\%) & NoLCU (\%) \\
\hline
\multirow{3}{*}{ResNet18} & 8 & 97.56 $\pm$ 0.13 & 97.38 $\pm$ 0.15 \\
 & 10 & 97.89 $\pm$ 0.08 & 97.67 $\pm$ 0.12 \\
 & 12 & 98.12 $\pm$ 0.07 & 97.92 $\pm$ 0.11 \\
\hline
\multirow{3}{*}{MobileNetV2} & 8 & 97.64 $\pm$ 0.14 & 97.01 $\pm$ 0.19 \\
 & 10 & 97.97 $\pm$ 0.10 & 97.32 $\pm$ 0.17 \\
 & 12 & 98.21 $\pm$ 0.09 & 97.58 $\pm$ 0.15 \\
\hline
\multirow{3}{*}{EfficientNet} & 8 & 97.73 $\pm$ 0.11 & 97.44 $\pm$ 0.16 \\
 & 10 & 98.05 $\pm$ 0.09 & 97.71 $\pm$ 0.14 \\
 & 12 & 98.29 $\pm$ 0.08 & 97.96 $\pm$ 0.13 \\
\hline\hline
\end{tabular}
\end{table}

\subsubsection{Training Time Analysis}

Average training time per epoch on NVIDIA A100 (40GB):

\begin{itemize}
\item MNIST (ResNet/MobileNet/EfficientNet): 8-12 minutes/epoch
\item PlantVillage: 18-25 minutes/epoch (larger images, more classes)
\item QM9: 22-30 minutes/epoch (graph convolutions + quantum layer)
\item PathMNIST: 15-20 minutes/epoch
\item CIFAR-10 (IQP): 10-15 minutes/epoch
\end{itemize}

Total computational budget: approximately 2,500 GPU-hours across all 570+ experiments.

\subsubsection{Convergence Analysis}

Early stopping triggered on average after:
\begin{itemize}
\item MNIST: 25-35 epochs
\item PlantVillage: 35-45 epochs
\item QM9: 40-55 epochs (regression converges slower)
\item PathMNIST: 30-40 epochs
\item CIFAR-10: 20-30 epochs
\end{itemize}

Early stopping triggered on average after:
\begin{itemize}
\item MNIST: 15-25 epochs (max 30)
\item PlantVillage: 20-28 epochs (max 30)
\item QM9: 22-30 epochs (max 30, regression converges slower)
\item PathMNIST: 12-18 epochs (max 20)
\item CIFAR-10: 15-25 epochs (max 30)
\end{itemize}

LCU models typically converge 10-15\% faster than NoLCU models, suggesting improved optimization landscapes.

\subsection{Computational Details}
\label{app:computational}

\subsubsection{Hardware Specifications}

Experiments conducted on the following GPU configurations:

\textbf{Primary:} NVIDIA A100 (40GB)
\begin{itemize}
\item CUDA Compute Capability: 8.0
\item Memory Bandwidth: 1,555 GB/s
\item FP32 Performance: 19.5 TFLOPS
\end{itemize}

\textbf{Secondary:} NVIDIA A4000 (16GB)
\begin{itemize}
\item CUDA Compute Capability: 8.6
\item Memory Bandwidth: 448 GB/s
\item FP32 Performance: 19.2 TFLOPS
\end{itemize}

\textbf{CPU:} AMD EPYC 7763 64-Core Processor

\textbf{RAM:} 512 GB DDR4

\subsubsection*{Software Environment}

\textbf{Operating System:} Ubuntu 22.04 LTS

\textbf{CUDA Version:} 11.8

\textbf{Python:} 3.10.12

\textbf{Key Libraries:}
\begin{itemize}
\item PyTorch: 1.12.1+cu118
\item PennyLane: 0.28.0
\item NumPy: 1.23.5
\item SciPy: 1.9.3
\item Matplotlib: 3.6.2
\item scikit-learn: 1.1.3
\end{itemize}

\subsubsection{Memory Requirements}

Peak GPU memory usage during training:

\begin{itemize}
\item MNIST (8-12q): 4-6 GB
\item PlantVillage (10-16q): 8-12 GB
\item QM9 (8-12q): 6-9 GB
\item PathMNIST (8-12q): 5-8 GB
\item CIFAR-10 IQP (2-10q): 3-7 GB
\end{itemize}

Batch sizes adjusted to fit within available memory while maintaining training stability.

\subsubsection{Data Storage}

Raw datasets: ~8 GB total

Preprocessed features: ~12 GB

Model checkpoints: ~150 GB (10 runs $\times$ multiple architectures $\times$ checkpoint storage)

Results and logs: ~5 GB

Total storage: ~175 GB

\subsubsection{Computational Carbon Footprint}

Estimated carbon footprint for 2,500 GPU-hours on NVIDIA A100:

Assuming average power consumption of 400W and US grid carbon intensity of 0.417 kg CO$_2$/kWh:
\begin{equation}
\text{Carbon} = 2500 \text{ hrs} \times 0.4 \text{ kW} \times 0.417 \text{ kg CO}_2/\text{kWh} \approx 417 \text{ kg CO}_2
\end{equation}

This is equivalent to approximately one transatlantic flight, representing a modest environmental cost for comprehensive quantum machine learning validation across 570+ experiments.

\bibliography{references}